\documentclass[12pt]{article}
\usepackage{amsmath}
\usepackage{amssymb}
\usepackage{epsfig}
\usepackage{graphicx}
\usepackage{cite}
\usepackage{multirow}
\usepackage{longtable}
\usepackage{lscape}
\usepackage{bbm}
\usepackage{mathrsfs}

\allowdisplaybreaks[4] % for page break

\newcommand{\capdef}{}
\newcommand{\mycaption}[2][\capdef]{\renewcommand{\capdef}{#2}%
        \caption[#1]{{\footnotesize #2}}}
\makeatletter
\renewcommand{\fnum@table}{\textbf{\tablename~\thetable}}
\renewcommand{\fnum@figure}{\textbf{\figurename~\thefigure}}
\makeatother

\newcounter{myenumi}

\renewcommand{\themyenumi}{\roman{myenumi}}
{\end{list}}

\newlength{\myem}
\newcounter{mysubequation}[equation]

\makeatletter
\renewcommand{\section}{\@startsection{section}{1}{0em}{-\baselineskip}%
{\baselineskip}{\normalfont\large\bfseries}}
\renewcommand{\subsection}%
{\@startsection{subsection}{2}{0em}{-0.7\baselineskip}%
{0.7\baselineskip}{\normalfont\bfseries}}
\makeatother

\newcommand{\bi}{\begin{itemize}}
\newcommand{\ei}{\end{itemize}}

\newcommand{\be}{\begin{equation}}
\newcommand{\ee}{\end{equation}}
\newcommand{\bea}{\begin{eqnarray}}
\newcommand{\eea}{\end{eqnarray}}

\newcommand{\ldm}{\Delta m_{31}^2}
\newcommand{\sdm}{\Delta m_{21}^2}
\newcommand{\deltacp}{\delta_{\mathrm{CP}}}
\newcommand{\stheta}{\sin^2 2 \theta_{13}}

\newcommand{\ie}{{\it i.e.}}

\newcommand{\eg}{{\it e.g.}}

\newcommand{\cf}{{\it cf.}}

\newcommand{\fig}{Fig.}
\newcommand{\Fig}{Fig.}

\newcommand{\Figs}{Figs.}
\newcommand{\Ref}{Ref.}
\newcommand{\Refs}{Refs.}
\newcommand{\Sec}{Sec.}

\newcommand{\Tab}{Table}

\newcommand{\figu}[1]{\fig~\ref{fig:#1}}

\begin{document}
%%%%%%%%%%%%%%%%%%%%%%%%%%%%%%%%%%%%%%%%%%%%%%%%%%%%%%%%%%%%%%%%%%%%%
%%%%                     Title-page                              %%%%
%%%%%%%%%%%%%%%%%%%%%%%%%%%%%%%%%%%%%%%%%%%%%%%%%%%%%%%%%%%%%%%%%%%%%

\begin{titlepage}

\renewcommand{\thefootnote}{\alph{footnote}}

%\vspace*{-3.cm}
\begin{flushright}
EURONU-WP6-09-06 \\
\end{flushright}

\vspace*{1cm}

\renewcommand{\thefootnote}{\fnsymbol{footnote}}
\setcounter{footnote}{-1}

{\begin{center}
{\large\bf
First hint for CP violation in neutrino oscillations from upcoming superbeam
and reactor experiments 
} \end{center}}
\renewcommand{\thefootnote}{\alph{footnote}}

\vspace*{.5cm}
{\begin{center} {\large{\sc
 		Patrick~Huber\footnote[1]{\makebox[1.cm]{Email:}
                pahuber$\cdot$at$\cdot$vt.edu},
 		Manfred~Lindner\footnote[2]{\makebox[1.cm]{Email:}
                lindner$\cdot$at$\cdot$mpi-hd.mpg.de}, \\
 		Thomas~Schwetz\footnote[3]{\makebox[1.cm]{Email:}
                schwetz$\cdot$at$\cdot$mpi-hd.mpg.de}, and
                Walter~Winter\footnote[4]{\makebox[1.cm]{Email:}
                winter$\cdot$at$\cdot$physik.uni-wuerzburg.de}
                }}
\end{center}}
\vspace*{0cm}
{\it
\begin{center}

\footnotemark[1]
	Department of Physics, IPNAS, Virginia Tech, Blacksburg, VA 24061, USA

\footnotemark[2]${}^,$\footnotemark[3]
	Max-Planck-Institut f{\"u}r Kernphysik, Postfach 103980, D-69029 Heidelberg, Germany

\footnotemark[4]
       Institut f{\"u}r Theoretische Physik und Astrophysik, Universit{\"a}t W{\"u}rzburg, \\
       D-97074 W{\"u}rzburg, Germany

\end{center}}

\vspace*{0.5cm}

{\large \bf
\begin{center} Abstract \end{center}  }

We compare the physics potential of the upcoming neutrino oscillation
experiments Daya Bay, Double Chooz, NO$\nu$A, RENO, and T2K based on
their anticipated nominal luminosities and schedules. After discussing
the sensitivity to $\theta_{13}$ and the leading atmospheric
parameters, we demonstrate that leptonic CP violation will hardly be
measurable without upgrades of the T2K and NO$\nu$A proton drivers,
even if $\theta_{13}$ is large. In the presence of the proton drivers,
the fast track to hints for CP violation requires communication
between the T2K and NO$\nu$A collaborations in terms of a mutual
synchronization of their neutrino-antineutrino run plans.  Even in
that case, upgrades will only discover CP violation in a relatively
small part of the parameter space at the $3\sigma$ confidence level,
while 90\% confidence level hints will most likely be obtained.
Therefore, we conclude that a new facility will be required if
the goal is to obtain a significant result with high probability.

\vspace*{.5cm}

\end{titlepage}

\newpage

\renewcommand{\thefootnote}{\arabic{footnote}}
\setcounter{footnote}{0}

%%%%%%%%%%%%%%%%%%%%%%%%%%%%%%%%%%%%%%%%%%%%%%%%%%%%%%%%%%%%%%%%%%%%%%%%%%%%%%%%%
\section{Introduction}

Neutrino oscillations have been firmly established in the last ten years or
so by a beautiful series of experiments with neutrinos from the
sun~\cite{Cleveland:1998nv, Altmann:2005ix, Hosaka:2005um, Ahmad:2002jz,
Aharmim:2008kc, Arpesella:2008mt}, the Earth's atmosphere~\cite{Fukuda:1998mi,
Ashie:2005ik}, nuclear reactors~\cite{Araki:2004mb, :2008ee}, and
accelerators~\cite{Ahn:2006zz, Adamson:2008zt}. While these measurements have
discovered and confirmed the dominant effective 2-flavor oscillation modes,
it will be the purpose of the upcoming generation of experiments to discover
sub-leading effects. This includes the
following tasks:
\begin{enumerate}
  \item
  Determination of the small lepton mixing angle $\theta_{13}$,
  \item
  Establishing CP violation (CPV) in neutrino oscillations for a 
  value of the Dirac CP phase $\deltacp \neq 0, \pi$,
  \item
  Identification of the type of the neutrino mass hierarchy (MH), which can
  be normal ($\ldm > 0$) or inverted ($\ldm < 0$).
\end{enumerate}

There are several neutrino oscillation experiments currently under
construction, which are expected to start data taking soon. These are the
reactor neutrino experiments Double Chooz~\cite{Ardellier:2004ui}, Daya
Bay~\cite{Guo:2007ug}, RENO~\cite{Kim:2008zzb} and the accelerator
experiments T2K~\cite{Itow:2001ee} and NO$\nu$A~\cite{Ambats:2004js}. The
primary goal for all of these experiments is the discovery of the yet
unknown mixing angle $\theta_{13}$. In this work, we
will revisit the nominal sensitivities of these projects to $\theta_{13}$
taking care of the different nature of the experiments ($\bar\nu_e$
disappearance in reactors versus $\nu_\mu\to\nu_e$ or
$\bar\nu_\mu\to\bar\nu_e$ appearance in accelerators), and estimate
the time evolution of the global discovery reach for $\theta_{13}$ based on
the official schedules of the experiments. This analysis updates previous
works~\cite{Minakata:2002jv, Huber:2002rs, Huber:2003pm, Huber:2004ug, McConnel:2004bd}
with respect to the now settled parameters for the considered experiments.
Let us mention that eventually already the currently running experiments
MINOS~\cite{Adamson:2008zt, minos-app} and OPERA~\cite{Rosa:2008zz} might
give a first hint for a non-zero value of $\theta_{13}$ in case it is close
to the present bound. We do not consider these experiments here, but focus
on the above facilities currently under preparation, since they will
clarify such hints with high significance.

Furthermore, we will study if  there is any chance to address items~2 and~3
above (CPV and MH) already with this set of upcoming experiments in case of a
soon coming positive signal for $\theta_{13}$, which implies
that $\theta_{13}$ is relatively ``large''. It turns out that even in this most
favorable case the sensitivity to CPV and MH of these experiments in their
nominal configuration is marginal. Therefore we will explore the
potential of ``minor upgrades'' to the proposed setups of T2K and NO$\nu$A,
based upon mostly existing equipment.
These include a longer running time and an upgraded beam power for both
experiments, and the addition of antineutrino running in T2K. Furthermore, we
investigate the optimization potential of a coordinated combination of T2K,
NO$\nu$A, and Daya Bay. The purpose of this analysis is to estimate whether
there will be any chance to have information on CPV or the MH around
2022--2025, given the rather optimistic assumptions concerning beam
performances and the size of $\theta_{13}$. 

Such considerations are relevant in view of the current international effort
towards a subsequent high precision neutrino facility~\cite{euronu, ids}.
These studies aim for a decision point concerning a future facility around
2012, where improved information on $\theta_{13}$ should be available from
the above mentioned experiments. If a non-zero value is
established until then, an eminent question will be wether CPV and MH 
can be found by upgrades of the existing
experiments, or indeed a new facility will be necessary. 

The outline of the paper is as follows. In Sec.~\ref{sec:experiments}, we
describe the planned experiments and give some details on our simulations. 
In Sec.~\ref{sec:nominal}, we consider the nominal configurations of the
experiments based on the information the experimental collaborations
provide. We study the sensitivity to $\theta_{13}$ (discovery potential as
well as the case of large $\theta_{13}$) and the improvements to be expected
in the leading atmospheric parameters $\theta_{23}$ and $|\Delta m^2_{31}|$.
In Sec.~\ref{sec:timescales}, we discuss the prospective time evolution of
the sensitivity to $\theta_{13}$ within the upcoming years. In
Sec.~\ref{sec:upgrades}, we consider possible upgrades for T2K and NO$\nu$A
focusing on the measurements of CPV and MH in the case of relatively large
$\theta_{13}$. Note that we arranged the Secs.~\ref{sec:nominal}
to~\ref{sec:upgrades} in an order from the most to the least established in
terms of the data provided by the experimental collaborations. Summary and
conclusions follow in Sec.~\ref{sec:conclusions}. In the technical appendix
we give details on our neutrino/antineutrino optimization algorithm.

%%%%%%%%%%%%%%%%%%%%%%%%%%%%%%%%%%%%%%%%%%%%%%%%%%%%%%%%%%%%%%%%%%%%%
\section{Experiment descriptions and simulation methods}
%%%%%%%%%%%%%%%%%%%%%%%%%%%%%%%%%%%%%%%%%%%%%%%%%%%%%%%%%%%%%%%%%%%%%
\label{sec:experiments}

Below we describe in some detail the considered experimental setups and our
simulation which we perform by using the GLoBES software~\cite{Huber:2004ka,Huber:2007ji}. 
The corresponding {\tt glb}-files are available at the GLoBES
web-page~\cite{Huber:2004ka} including detailed technical information on the simulation. 
In all cases our strategy is to follow as close as possible the original Letters of
Intent (LOIs) or Technical Design Reports (TDRs). We have made sure that our
sensitivities agree with the ``official'' curves from the corresponding
collaborations under the same assumptions. \Tab~\ref{tab:nom} summarizes the
key parameters of the considered experiments.

\begin{table}[t]
\begin{tabular}{lccrrlr}
\hline
Setup & $t_\nu$ [yr] & $t_{\bar{\nu}}$ [yr] & $P_{\mathrm{Th}}$ or
$P_{\mathrm{Target}}$ & $L$ [km] & Detector technology & $m_{\mathrm{Det}}$ \\
\hline
 Double Chooz & - & 3 &  8.6 GW & 1.05 & Liquid scintillator &  8.3 t\\
 Daya Bay     & - & 3 & 17.4 GW & 1.7  & Liquid scintillator &  80 t \\
 RENO         & - & 3 & 16.4 GW & 1.4  & Liquid scintillator & 15.4 t \\
 T2K          & 5 & - & 0.75~MW & 295  & Water Cerenkov      & 22.5 kt \\
 NO$\nu$A     & 3 & 3 & 0.7~MW  & 810  & TASD                &   15 kt \\
\hline
\end{tabular}
\mycaption{\label{tab:nom} Summary of the standard setups at their nominal luminosities.}
\end{table}

Reactor experiments look for the disappearance of $\bar\nu_e$,
governed by $\theta_{13}$, where the neutrinos are produced
in the nuclear fission processes in commercial nuclear power plants. 
All three experiments, Double Chooz, Daya
Bay, and RENO, use a liquid scintillator doped with Gadolinium in
order to exploit the coincidence of a positron and a neutron from the
reaction $\bar\nu_e + p \to e^+ + n$. The two crucial parameters
determining the final sensitivity are the total exposure (proportional
to the detector mass times the time-integrated thermal power of the
reactor over the lifetime of the experiment) and the systematic
uncertainty, see, \eg, \Ref~\cite{Huber:2003pm} for a discussion. All
three proposals use the concept of near and far detectors in order to
reduce uncertainties on the initial neutrino flux. Our implementation
of systematic uncertainties is similar to the one of
\Ref~\cite{Huber:2006vr}. While for Double Chooz and Daya Bay
detailed proposals are available, \Refs~\cite{Ardellier:2004ui}
and~\cite{Guo:2007ug}, respectively, the RENO experiment is somewhat
less documented and some properties have to be extrapolated from the
other two proposals. The characteristics of the three reactor
experiments are (see also \Ref~\cite{Mention:2007um} for a comparison
study):
\begin{description}
\item[Double Chooz:] The Chooz power plant in France consists of two
  reactors with 4.3~GW thermal power each. There will be two
  detectors with 8.3~t fiducial mass each, a far detector at a distance
  of 1.0~km and 1.1~km from the two cores, and a near detector at a
 distance of 470~m and
  350~m, respectively. Including efficiencies of 80\%, detector dead times, and a
  load factor of 78\% for the reactors, the event rates per year are
  $8\times 10^4$ and $1.5\times 10^4$ for the near and far detectors,
  respectively. The uncertainty on the relative normalization of the
  detectors is assumed to be 0.6\%.
\item[Daya Bay:] The Chinese Daya Bay reactor complex consists
  currently of two pairs of reactor cores (Daya Bay and Ling Ao),
  separated by about 1.1~km. The complex generates 11.6~GW of thermal
  power; this will increase to 17.4~GW by early 2011 when a third pair
  of reactor cores (Ling Ao II) is put into operation. For the 3 years
  nominal exposure we assume that all the mentioned six cores are
  operational.  In total, eight detector modules will be installed,
  20~t fiducial mass each. The far site consists of 4 modules with
  distances from the three reactors of 1.985, 1.613, and 1.618~km.
  There will also be two near labs, each containing two detector
  modules at distances from the reactors between 360 and 530~m. The
  event rate per year is about $4\times 10^5$ at the near sites and
  about $10^5$ at the far site. The detector normalization is assumed
  to be 0.18\% uncorrelated between the detector modules. This
  corresponds to the final goal in terms of systematics without
  swaping detector modules.
\item[RENO:] The RENO experiment will be located at the Yonggwang
  reactor complex in South Korea, which consists of 6 cores equally
  distributed along a straight line of about 1.5~km, with a total
  thermal power of 16.4~GW. Two detectors with about 15~t fiducial
  mass each will be installed roughly at distances of 320~m and 1.4~km
  from the reactor line. Event rates per year will be about $6\times
  10^5$ at the near site and about $3\times 10^4$ at the far site. The
  relative detector normalization is assumed at 0.5\%.
\end{description}

The T2K and NO$\nu$A superbeam experiments search for the appearance
of electron neutrinos in a beam of mainly muon neutrinos, from the
decay of pions and kaons produced at a proton accelerator. Such beams
unavoidably contain some intrinsic electron neutrinos which consist a
background for the appearance search. Both experiments explore the
off-axis technique to suppress the electron neutrino as well as the
neutral current backgrounds and to achieve a more peaked beam
spectrum. For both experiments, we also include the disappearance channels.
\begin{description}
\item[T2K:] The T2K experiment~\cite{Itow:2001ee} sends a neutrino beam from
  the J-PARC accelerator to the Super-Kamiokande water Cerenkov detector
  with a fiducial mass of 22.5~kt at a distance of 295~km. Our simulation is
  based on publicly accessible sources as of 2008. We assume a 2.5 degree
  off-axis beam corresponding to 0.75~MW of beam power for 5 years of
  neutrino running. The analysis of the $\nu_e$ appearance follows the
  thesis of M.~Fechner~\cite{Fechner2006}. Event rates and spectra for total
  signal and total background match \Fig~6.8 of \Ref~\cite{Fechner2006}, and
  systematics are taken from chapter~6 of that reference. Our
  sensitivity from the appearance channel agrees with \Fig~6.17 of
  \Ref~\cite{Fechner2006}. The disappearance analysis is taken from a talk
  by I.~Kato given at Neutrino 2008~\cite{Kato:2008zz}. From that source we
  obtain rates and spectra for signal and non-QE background.
\item[NO$\boldsymbol{\nu}$A:] The description of NO$\nu$A concerning the
  $\nu_e / \bar\nu_e$ appearance signals follows the proposal as of March
  15, 2005~\cite{Ambats:2004js}, the description of the disappearance signal
  is taken from \Ref~\cite{Yang_2004}. We calibrated our event simulation to
  the numbers for signal, background, and efficiencies given in the October
  2007 TDR~\cite{nova_TDR} (Tabs.~6.2--6.4). NO$\nu$A is a Totally Active
  Scintillator Detector (TASD) with a mass of 15~kt and is located at a
  distance of 810~km from NUMI beam source at Fermilab. The nominal
  luminosity is $18\times 10^{20}$ protons on target for neutrinos and
  antineutrinos, each. This number of protons on target is assumed to
  correspond to 3 years running at 0.7~MW beam power. While this equal
  exposure to neutrinos and antineutrinos is the default assumption, we
  will also consider the sensitivities in case of neutrino running only
  in some of the next sections. We
  adopt systematical errors of 5\% on the signal (not stated in the TDR) and
  10\% on the background (as given in the TDR). We use an analysis energy
  window from 1 to 3~GeV and assume the so-called medium energy (ME) beam
  configuration. 
\end{description}

For the sensitivity analyses we use the oscillation parameter values
from \Ref~\cite{Schwetz:2008er}: $\sdm=7.65 \cdot 10^{-5} \,
\mathrm{eV}^2$, $|\ldm|=2.40 \cdot 10^{-5} \, \mathrm{eV}^2$, $\sin^2
\theta_{12}=0.304$, and $\sin^2 \theta_{23}=0.500$, unless stated
otherwise. We impose external $1\sigma$ errors on $\sdm$ (3\%) and
$\theta_{12}$ (4\%) as conservative estimates for the current
measurement errors~\cite{Schwetz:2008er}, as well as $\ldm$ (5\%) for
reactor experiments if analyzed without beam experiments. In addition,
we include a 2\% matter density uncertainty.

%%%%%%%%%%%%%%%%%%%%%%%%%%%%%%%%%%%%%%%%%%%%%%%%%%%%%%%%%%%%%%%%%%%%%
\section{Physics potential at nominal luminosities}
%%%%%%%%%%%%%%%%%%%%%%%%%%%%%%%%%%%%%%%%%%%%%%%%%%%%%%%%%%%%%%%%%%%%%
\label{sec:nominal}

In this section, we discuss the physics potential of the experiments listed
in \Tab~\ref{tab:nom} based on their nominal luminosities, \ie, the running
times, detectors masses, and total numbers of protons (or thermal reactor
powers) anticipated by the respective collaborations.

\subsection{Discovery potentials for $\boldsymbol{\theta_{13}}$, MH, and CPV}

\begin{figure}[t]
\begin{center}
\includegraphics[width=\textwidth]{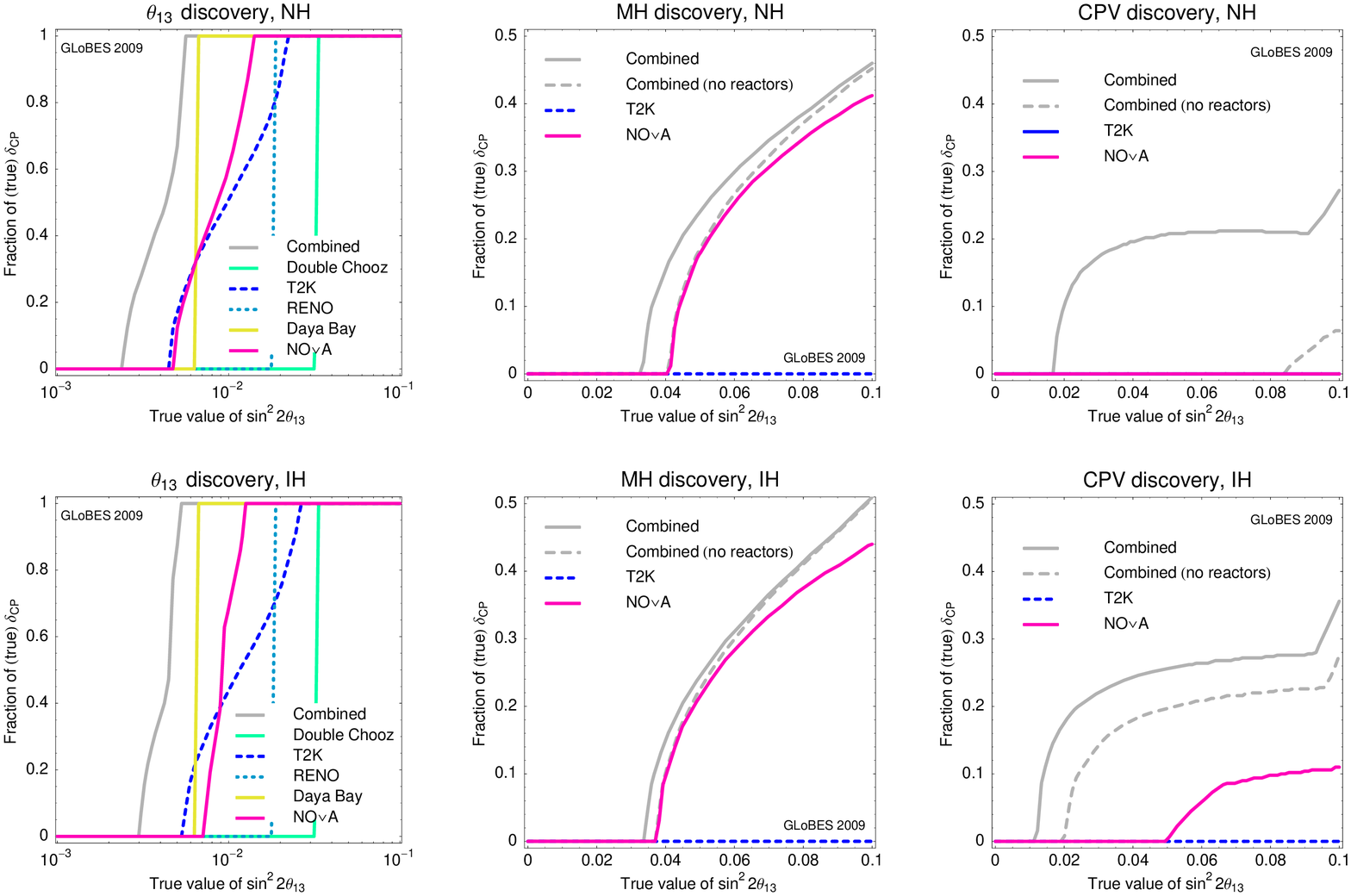}
\end{center}
\mycaption{\label{fig:disc} $\theta_{13}$, MH, and CPV discovery potential
  as fraction of true $\deltacp$ as a function of the true $\stheta$ for the
  normal hierarchy (upper row) and inverted hierarchy (lower row) at the
  90\% CL. Note the different vertical scales in the different panels.}
\end{figure}

We show the discovery potentials for $\theta_{13}$, MH, and CPV in
\figu{disc}. These discovery potentials quantify for any given (true)
$\stheta$ for which fraction of possible (true) values of $\deltacp$ the
corresponding quantity will be discovered.\footnote{In this work we use the
terms {\it true value} and {\it simulated value} synonymously in order to
denote the values of parameters adopted to generate ``data'', in contrast to
the {\it fit values} used to fit these data.} For $\theta_{13}$ this means
excluding the value $\theta_{13}=0$. The mass hierarchy discovery potential
is defined as the ability to exclude any degenerate solution for the wrong
(fit) hierarchy at the chosen confidence level. Similarly, the CP violation
discovery potential refers to the ability to exclude the CP conserving
solutions $\deltacp=0$ and $\deltacp=\pi$ for any degenerate solution at the
chosen confidence level. These discovery potentials of course depend on the
true values of $\theta_{13}$ and $\deltacp$ and the true hierarchy. In
\figu{disc} we show for a given true value of $\stheta$ (horizontal axis)
and a given true hierarchy (upper row normal, lower row inverted) the
fraction of all possible true values of $\deltacp$ for which the discovery
can be achieved at the 90\% confidence level. Hence, a fraction of
$\deltacp$ of unity (or 100\%) for a given $\stheta$ corresponds to a
discovery for any possible value of $\deltacp$.

The $\theta_{13}$ discovery potential (\cf, left panels of
\figu{disc}) of the reactor experiments does not depend on $\deltacp$
since by convention this phase does not appear in the disappearance
probability $P_{ee}$.  Furthermore, the probability is given to good
approximation by an effective 2-flavor expression:
$P_{ee}^\mathrm{react} \approx 1- \stheta \sin^2(\ldm L/4 E)$. Thanks to
the large exposure, Daya Bay will have the best discovery potential
among the reactor experiments of $\stheta = 0.0066$ at the 90\%~CL,
compared to 0.018 for RENO and 0.033 for Double Chooz.\footnote{Let us
  mention that the Daya Bay assumptions of a systematical error of
  0.18\%, fully uncorrelated among all detectors is more aggressive
  than for other reactor experiments.  For example, if the
  systematic error is at the level of 0.6\%, such as assumed in Double
  Chooz, the Daya Bay sensitivity of $\sin^22\theta_{13}= 0.0066$
  deteriorates to $\sin^22\theta_{13}\simeq 0.01$. If on the other hand the
  systematic error is $0.38\%$ and assumed to be fully correlated
  among modules at one site the limit would
  $\sin^22\theta_{13}\simeq0.012$~\cite{McFarlane:2009xx}. See also the
  discussion in \Ref~\cite{Mention:2007um}.}
In contrast, the $\nu_\mu\to\nu_e$ appearance probability relevant for
the beam experiments shows a dependence on the CP phase due to an
interference term of the oscillations with $\ldm$ and $\sdm$. A
discussion of the complementarity of reactor and beam experiments can
be found, \eg, in \Refs~\cite{Huber:2003pm, Minakata:2002jv,
McConnel:2004bd}.
Whether the best $\theta_{13}$ discovery potential is obtained from
Daya Bay or from one of the beams depends on the true $\deltacp$. For
instance, for the normal hierarchy, T2K and NO$\nu$A will have a
discovery potential for slightly smaller values of $\stheta$ than Daya Bay for
about 30\% of all possible $\deltacp$ values.

The mass hierarchy measurement (\cf, middle panels of \figu{disc})
requires NO$\nu$A because of its relatively long baseline and
therefore significant matter effects. T2K does not have any mass
hierarchy discovery potential.  For very large $\stheta$, the mass
hierarchy will be discovered for about 40--50\% of all values of
$\deltacp$ at 90\%~CL. There is no sensitivity left at this CL for
$\stheta < 0.04$. Only minor improvement can be achieved if other
experiments are added to NO$\nu$A; the combination with T2K helps
somewhat for large $\theta_{13}$, whereas the reactor experiments
contribute somewhat for small $\theta_{13}$.

For the discovery of CP violation (\cf, right panels of \figu{disc}),
neither T2K nor NO$\nu$A alone have a substantial potential. Only the
combination of these two experiments can measure CPV at 90\%~CL for up to
30\% of all values of $\deltacp$ if the hierarchy is inverted. This
hierarchy dependence appears because the matter effect in NO$\nu$A leads to
a more balanced neutrino-antineutrino statistics for the inverted hierarchy
due to an matter enhancement of the antineutrino event rate. In the
presence of the reactor experiments, however, the CP violation discovery
potential improves significantly, especially for the normal hierarchy. If the
hierarchy is normal, Daya Bay may in fact be the key prerequisite to obtain
an early hint for CP violation.

We stress that these sensitivities for MH and CPV are at the 90\%~CL,
which by no means can be considered as a discovery. Increasing the CL
leads to drastic loss in sensitivity and almost nothing can be said about CPV
and MH at the 3$\sigma$ CL.

%%%%%%%%%%%%%%%%%%%%%%%%%%%%%%%%%%%%%%%%%%%%%%%%%%%%%%%%%%%%%%%%%%%%%%%%%%%%%%%%
\subsection{The case of $\boldsymbol{\theta_{13}}$ just around the corner}

\begin{figure}[p!]
\begin{center}
\includegraphics[width=\textwidth]{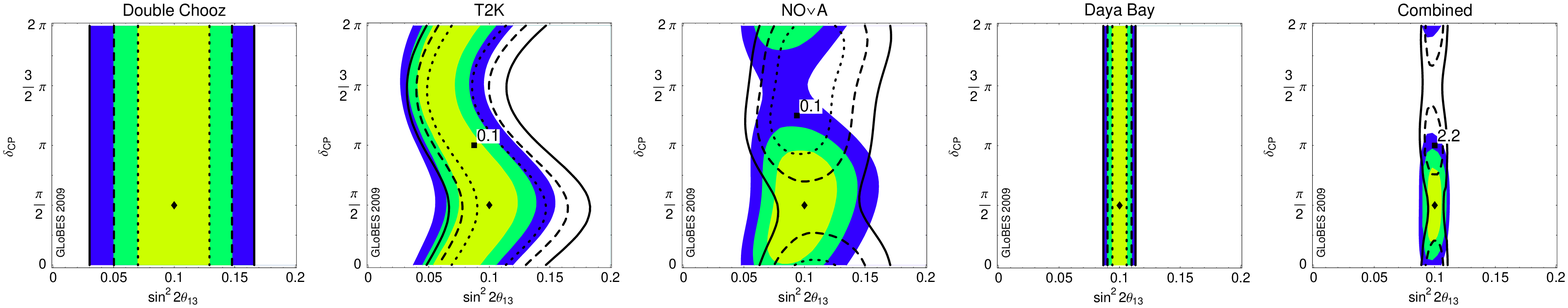} \\
\includegraphics[width=\textwidth]{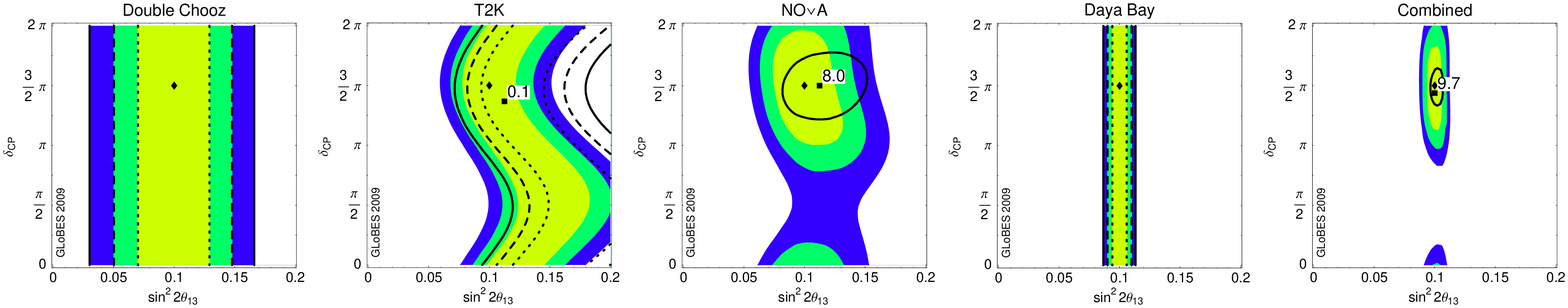}
\end{center}
  \mycaption{\label{fig:deltatheta} Fits in the $\theta_{13}$-$\deltacp$ plane
  for $\stheta=0.1$ and $\deltacp=\pi/2$ (upper row) and $\deltacp=3
  \pi/2$ (lower row). A normal simulated hierarchy is assumed. The contours
  refer to $1\sigma$, $2 \sigma$, and $3 \sigma$ (2 d.o.f.). The fit
  contours for the right fit hierarchy are shaded (colored), the ones for
  the wrong fit hierarchy fit are shown as curves. The best-fit values are
  marked by diamonds and boxes for the right and wrong hierarchy,
  respectively, where the minimum $\chi^2$ for the wrong hierarchy is
  explicitely shown.}
\end{figure}

\begin{figure}[p!]
\begin{center}
\includegraphics[width=\textwidth]{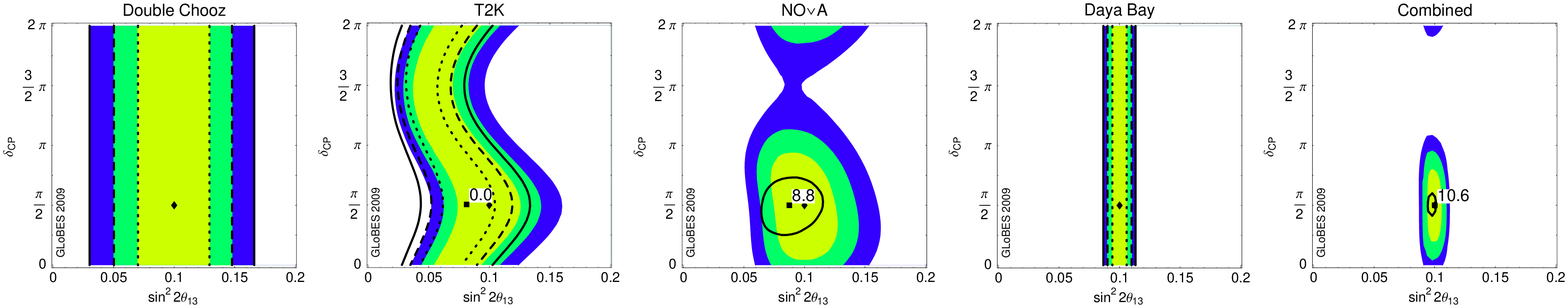} \\
\includegraphics[width=\textwidth]{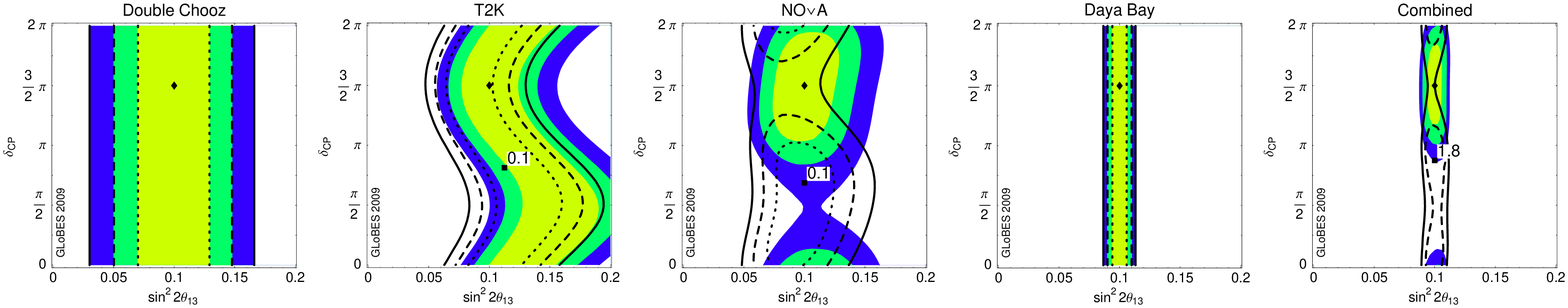}
\end{center}
\mycaption{\label{fig:deltathetai} Same as \figu{deltatheta}, for the
inverted simulated hierarchy.}
\end{figure}

We show in \figu{deltatheta} (normal simulated hierarchy) and
\figu{deltathetai} (inverted simulated hierarchy) how typical fits in the
$\theta_{13}$-$\deltacp$ plane would look like if $\theta_{13}$ was large
($\stheta=0.1$) and $\deltacp$ was close to maximal CP violation
$\deltacp=\pi/2$ (upper rows) and $\deltacp=3 \pi/2$ (lower rows). The fit
contours for the right fit hierarchy are shaded (colored), the ones for the
wrong fit hierarchy fit are shown as curves.\footnote{Here the ``right'' fit
hierarchy is the same as the simulated hierarchy, and the ``wrong'' fit
hierarchy is the other fit hierarchy.} 

The figures show the characteristics of the different classes of
experiments: The reactor experiments do not depend on $\deltacp$, and the
wrong fit hierarchy coincides with the right hierarchy. For T2K, which is
simulated with neutrino running only, there is some dependence on
$\deltacp$, but the correlation between $\deltacp$ and $\theta_{13}$ cannot
be resolved without antineutrino running. The wrong hierarchy contours are
slightly shifted, but the minimum $\chi^2$ is close to zero. NO$\nu$A, on
the other hand, has both neutrino and antineutrino running in our
simulation, which means that the correlation can, at least in principle, be
resolved. The wrong hierarchy can in some cases be excluded because of 
matter effects. In the combination of the experiments, the
combination between Daya Bay and the beams allows for a substantial
reduction of the allowed parameter space due to almost orthogonal
measurements. In the most optimistic cases, the mass hierarchy can be
determined at $3 \sigma$ confidence level, and maximal CP violation can
be established at relatively modest confidence as well. However, note that
these optimistic cases represent only a very small fraction of the parameter
space.

%%%%%%%%%%%%%%%%%%%%%%%%%%%%%%%%%%%%%%%%%%%%%%%%%%%%%%%%%%%%%%%%
\subsection{Leading atmospheric parameters}

\begin{figure}[t]
\begin{center}
\includegraphics[width=\textwidth]{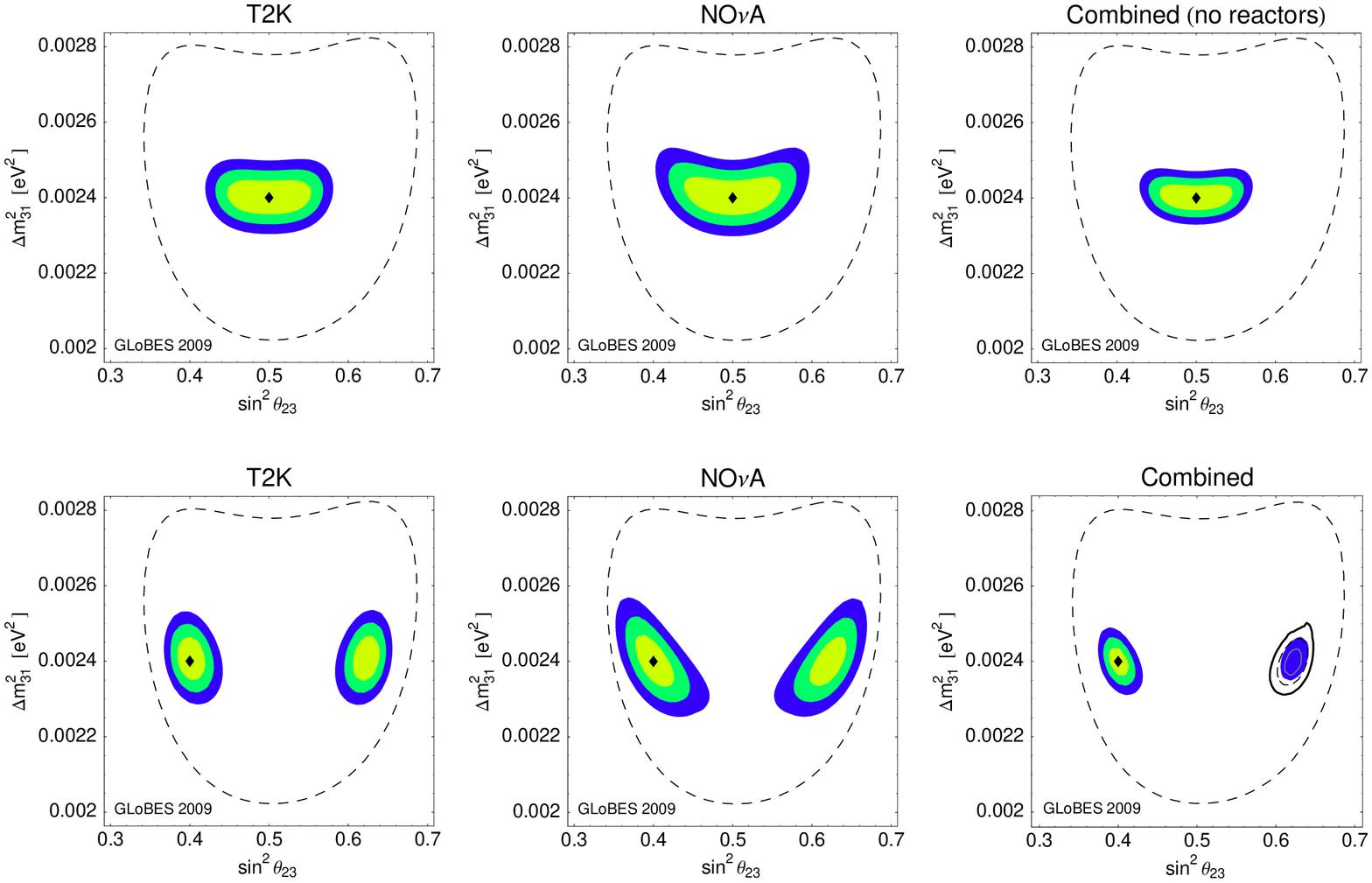}
\end{center}
\mycaption{\label{fig:atm} Fits in $\sin^2 \theta_{23}$-$\ldm$ plane
  for two different sets of simulated values: $\sin^2
  \theta_{23}=0.5$, $\stheta=0$ (upper row) and $\sin^2
  \theta_{23}=0.4$, $\stheta=0.1$, $\deltacp=0$ (lower row).  The
  upper right panel does not change significantly if the reactor
  experiments are added. In the lower right panel, the unshaded
  contours are for the combination of all experiments without reactor
  experiments. The currently allowed MINOS+atmospheric region is shown
  as dashed curves in all panels ($3 \sigma$ confidence
  level)~\cite{Schwetz:2008er}. The contours corresponds to the $1
  \sigma$, $2 \sigma$, and $3 \sigma$ confidence level, respectively
  (2 d.o.f.).  }
\end{figure}

No matter if $\theta_{13}$ will be discovered or not, the upcoming
beam experiments will allow for precision measurements of the leading
atmospheric parameters by exploring the $\nu_\mu$ disappearance
channel. We illustrate the improvement of the currently allowed
parameter range (dashed curves) in the $\sin^2 \theta_{23}$-$\ldm$
plane in \figu{atm} for the two different scenarios $\sin^2
\theta_{23}=0.5$, $\stheta=0$ (upper row) and $\sin^2
\theta_{23}=0.4$, $\stheta=0.1$, $\deltacp=0$ (lower row). In the
maximal mixing case, the allowed region can be substantially reduced
by the beam experiments, especially the $\ldm$ interval, with hardly
any dependence on the reactor experiments or the value of
$\theta_{13}$. 

In the lower row of \figu{atm}, a relatively large deviation from
maximal atmospheric mixing is chosen. Again, the allowed parameter
ranges can be significantly reduced by the beams, apart from an octant
ambiguity in $\theta_{23}$~\cite{Fogli:1996pv}.  As illustrated in the
lower right panel this ambiguity might be resolved if $\theta_{13}$ is
large and beams are combined with an accurate reactor experiment,
compare the shaded (including reactors) and unshaded (excluding
reactors) regions. In this case the two solutions corresponding to the
two $\theta_{23}$-octants for the appearance channel of the beams are
located at rather different values of $\theta_{13}$. Therefore, an
independent determination of $\theta_{13}$ from reactors can in
principle resolve the ambiguity~\cite{Minakata:2002jv,
McConnel:2004bd}.

\begin{figure}[t]
\begin{center}
\includegraphics[width=0.75\textwidth]{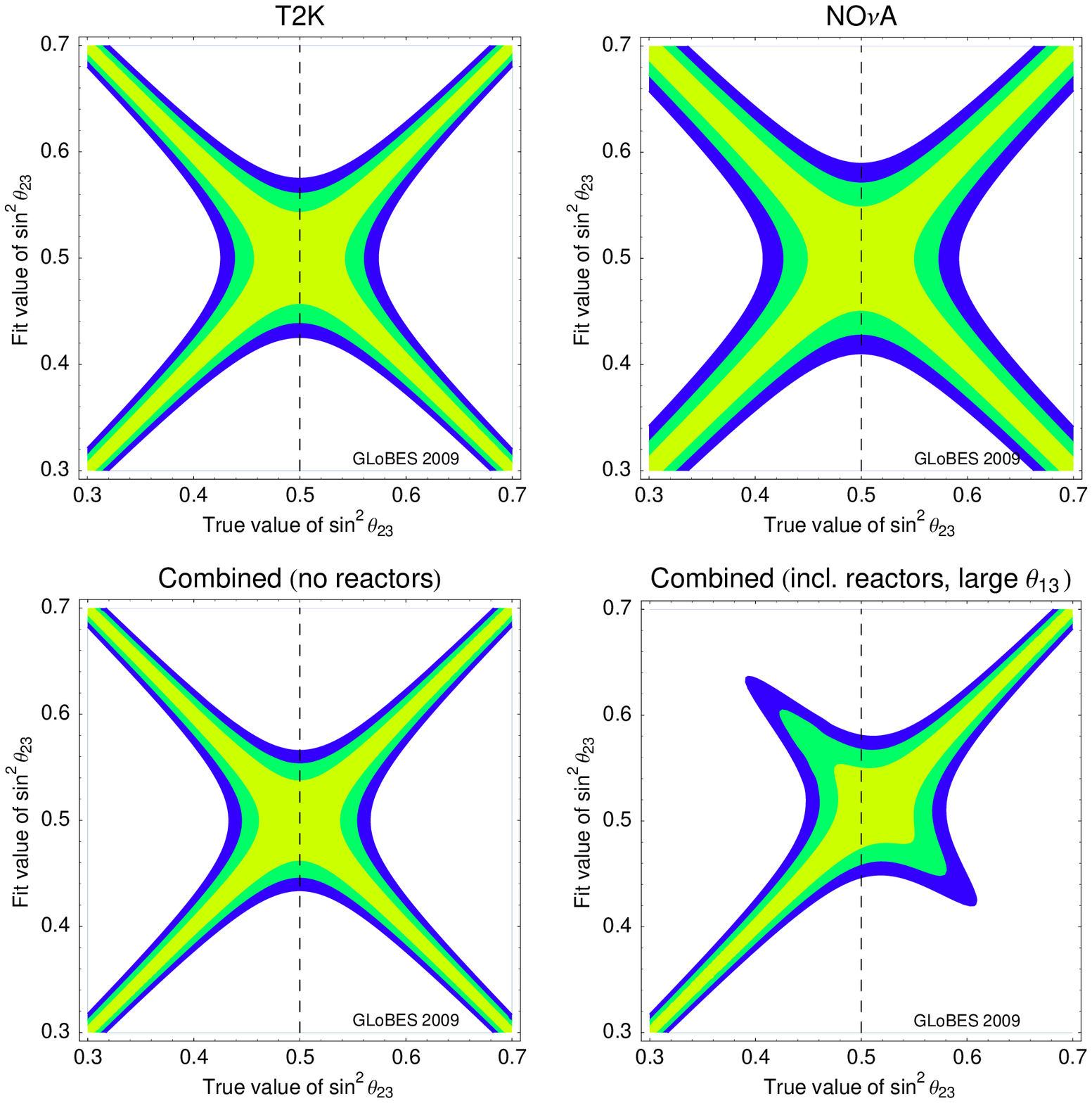}
\end{center}
\mycaption{\label{fig:theta23} Fit range in $\theta_{23}$ as a
  function of the true $\theta_{23}$ for different experiments
  ($1\sigma$, $2\sigma$, $3 \sigma$ for 1~d.o.f.).  The upper left and
  right, and the lower left panels are computed for $\stheta=0$,
  whereas the lower right panel is for $\stheta=0.1$,
  $\deltacp=0$. The qualitative picture is hardly affected by the
  reactor experiments or large $\theta_{13}$, unless both conditions
  apply (large $\theta_{13}$ and reactor experiments).  }
\end{figure}

In order to discuss the sensitivity of deviations from maximal
atmospheric mixing and the octant resolution potential as a function
of the true $\theta_{23}$, we show in \figu{theta23} the fit range in
$\theta_{23}$ as a function of the true $\theta_{23}$ for different
experiments. 
Deviations from maximal atmospheric mixing are theoretically
interesting because they may point towards deviations from a symmetry,
see \Ref~\cite{Antusch:2004yx} for a more detailed discussion. In
\figu{theta23}, such a deviation can be established if $\sin^2
\theta_{23}=0.5$ can be excluded at the vertical axis. Obviously, this
measurement does not require the presence of the reactor experiments
or large $\theta_{13}$.  The combination of the beams can establish
deviations from maximal atmospheric mixing for $|\sin^2 \theta_{23} -
0.5| \gtrsim 0.07$ ($3 \sigma$).

The sensitivity to the octant of $\theta_{23}$ can be interesting to
distinguish certain models. For instance, sum rules such as $\theta_{23}
\simeq \pi/4 \pm \theta_{13}^2/2$ might be tested, where the sign depends on
the model assumptions~\cite{Niehage:2008sg}. If deviations from maximality
are due to renormalization group effects, in a large class of models the
sign of the deviation is different for normal and inverted hierarchy, and
depends on the presence of Supersymmetry~\cite{Antusch:2003kp}. As
illustrated by the lower two panels in \figu{theta23}, both the combination
with the reactor data and a large $\theta_{13}$ are necessary to perform
this measurement with the given experiments. Sensitivity to the octant of
$\theta_{23}$ is present if any $\theta_{23}$ in the wrong octant can be
excluded. There is no octant sensitivity in the upper and lower left panels
of \figu{theta23}. From the lower right panel, we can read off octant
sensitivity if $\sin^2 \theta_{23} \lesssim 0.39$ or $\sin^2 \theta_{23}
\gtrsim 0.61$ ($3 \sigma$, $\stheta=0.1$).

%%%%%%%%%%%%%%%%%%%%%%%%%%%%%%%%%%%%%%%%%%%%%%%%%%%%%%%%%%%%%%%%%%%%%%%%%%%%%%%%%%%%%
\section{Sensitivity time evolution}
%%%%%%%%%%%%%%%%%%%%%%%%%%%%%%%%%%%%%%%%%%%%%%%%%%%%%%%%%%%%%%%%%%%%%%%%%%%%%%%%%%%%%
\label{sec:timescales}

In the following, we discuss the sensitivity evolution within the coming
years for different experiments, based as much as possible on official
statements of the collaborations. Although the assumed schedules and proton
beam plans may turn out to be not realistic in some cases, our toy scenario
will be illustrative to show the key issues for the individual experiments
within the global neutrino oscillation context.  We will show the
sensitivities as a function of time assuming that data are continously
analyzed and results are available immediately.  The key assumptions for our
toy scenario are as follows.

\begin{description}
\item[Double Chooz] starts 09/2009 and runs 1.5 years with far detector only,
  then with far and near detector~\cite{REFDC}. We assume that the
  experiment ends after five years. 
\item[Daya Bay] Starts 07/2011. At this time, it is assumed that 
  near and far detector halls are completed, and all modules are
  ready~\cite{REFDYB}. Furthermore, all three
  pairs of reactor cores are assumed to be online. Again, we limit the
  operation time to five years. 
\item[RENO] starts 06/2010~\cite{REFRENO} with both detectors and runs for 5 years. 
\item[T2K] starts 09/2009 with virtually 0~MW beam power, which increases linear to
  0.75~MW reached in 12/2012. From then we assume the full
  target power of 0.75~MW. (This is an approximation for the T2K proton plan
  from \Ref~\cite{REFT2K}). The Super-Kamiokande detector is online from the
  very beginning and the beam runs with neutrinos only, (at least) until
  2018 or 2019.
\item[NO$\nu$A] starts 08/2012 with full beam (0.7~MW), but 2.5~kt detector 
  mass only.  Then the detector mass increases linearly to 15~kt in 01/2014.
  From then we assume the full detector mass of 15~kt~\cite{REFNOVA}. The
  beam runs with neutrinos first, until the equivalent of three years
  operation at nominal luminosity (\cf, \Tab~\ref{tab:nom}) is reached, \ie,
  03/2016. Then it switches (possibly) to antineutrinos and runs at least
  until 2019.
\end{description}
%

%%%%%%%%%%%%%%%%%%%%%%%%%%%%%%%%%%%%%%%%%%%%%%%%%%%%%%%%%%%%%%%%%%%%%%%%%%%%%%%%%%%%%%%%%%%%
\subsection{Finding versus constraining $\boldsymbol{\theta_{13}}$}

We now discuss the sensitivity of the different experiments to
$\theta_{13}$. We include two qualitatively different aspects in the
discussion: The $\theta_{13}$ sensitivity limit and the $\theta_{13}$
discovery potential. The $\theta_{13}$ sensitivity limit describes the
ability of an experiment to constrain $\theta_{13}$ if no signal is seen. It
is basically determined by the worst case parameter combination which may
fake the simulated $\theta_{13}=0$. 
The sensitivity limit does not depend on the simulated hierarchy and
$\deltacp$, as the simulated $\theta_{13}=0$. For a more detailed
discussion, see \Ref~\cite{Huber:2004ug}, App.~C. The $\theta_{13}$
discovery potential is given by the smallest true value of $\theta_{13} > 0$
which cannot be fitted with $\theta_{13}=0$ at a given CL. Since the
simulated $\theta_{13}$, $\deltacp$, and hierarchy determine the simulated
rates, the $\theta_{13}$ discovery potential will depend on the values
of all these parameters chosen by nature. On the other hand, correlations
and degeneracies are of minor importance because for the fit $\theta_{13}=0$
is used.
The smallest $\theta_{13}$ discovery potential for all values of $\deltacp$
and the MH (risk-minimized $\theta_{13}$ discovery potential) is often
similar to the $\theta_{13}$ sensitivity limit. This holds to very good
approximation for reactor experiments, where statistics are Gaussian and the
oscillation physics is simple. For beam experiments differences occur due to
Poisson statistics as well as more complicated oscillation physics implying
correlations and degeneracies. 

\begin{figure}[t]
\begin{center}
\includegraphics[width=0.6\textwidth]{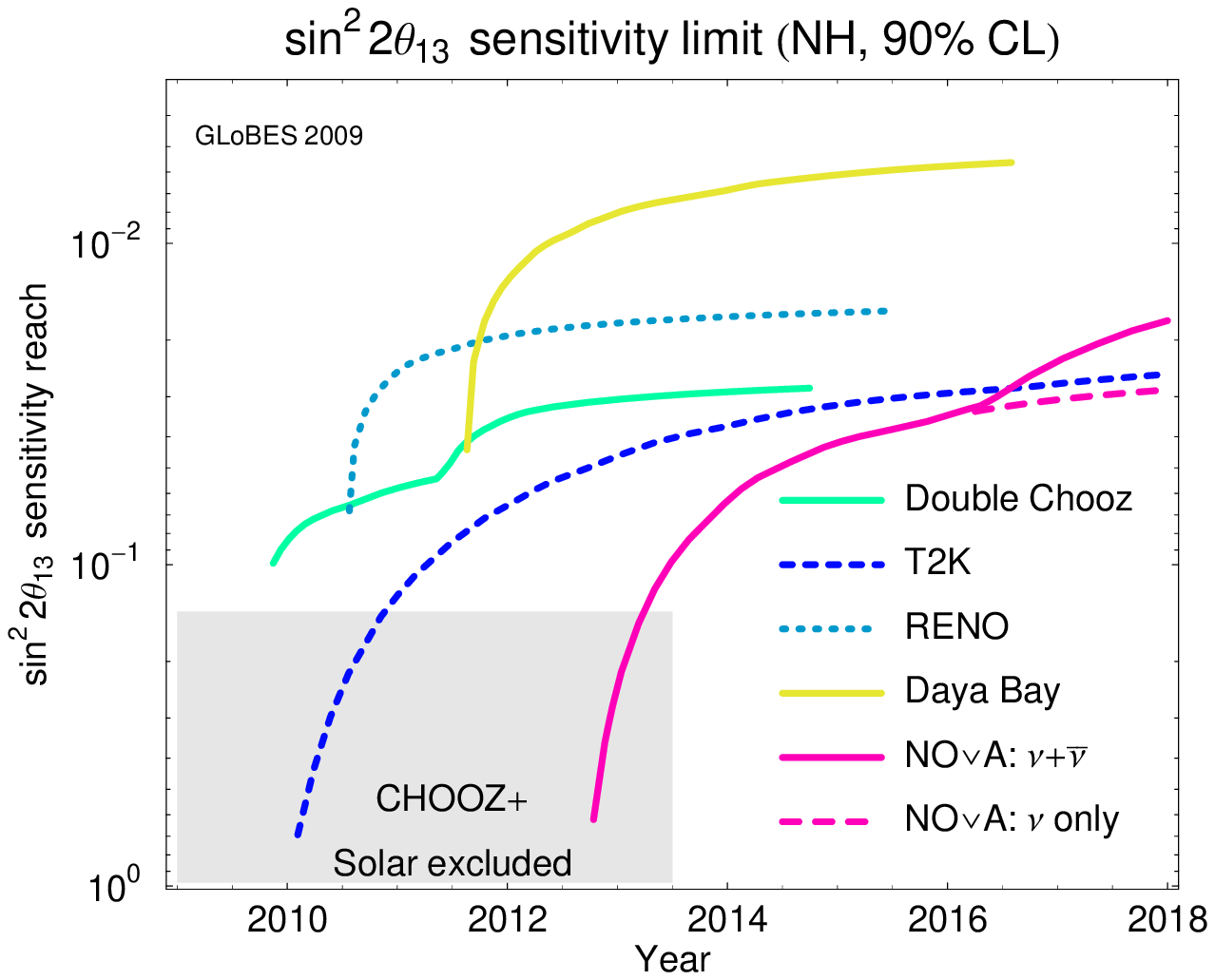}
\end{center}
\mycaption{\label{fig:evolsens} Evolution of the $\theta_{13}$ sensitivity
limit as a function of time (90\% CL), \ie, the 90\%~CL limit which will be
obtained if the true $\theta_{13}$ is zero.}
\end{figure}

We show the $\theta_{13}$ sensitivity limit as a function of time in
\figu{evolsens}. We observe that the global sensitivity limit will be
dominated by reactor experiments. As soon as operational, Daya Bay will
dominate the global limit. For Daya Bay, time is not critical, but matching
the systematics or statistics goals is. If
the assumed schedules of both, Double Chooz and Daya Bay are matched, Double
Chooz will dominate the $\theta_{13}$ sensitivity for about two years in the
absence of RENO. If available, RENO, on the other hand, will dominate the
$\theta_{13}$ sensitivity if it is operational significantly before the end
of 2011. Since we do not obtain any CP violation or mass hierarchy
sensitivity before 2014, as we shall demonstrate later, the reactor
contribution to those will be completely dominated by Daya Bay. As a
peculiarity, the $\theta_{13}$ sensitivity of NO$\nu$A is improved by
switching to antineutrinos. However, the global limit will at that time be
dominated by the reactor experiments.

\begin{figure}[t!]
\begin{center}
\includegraphics[width=0.48\textwidth]{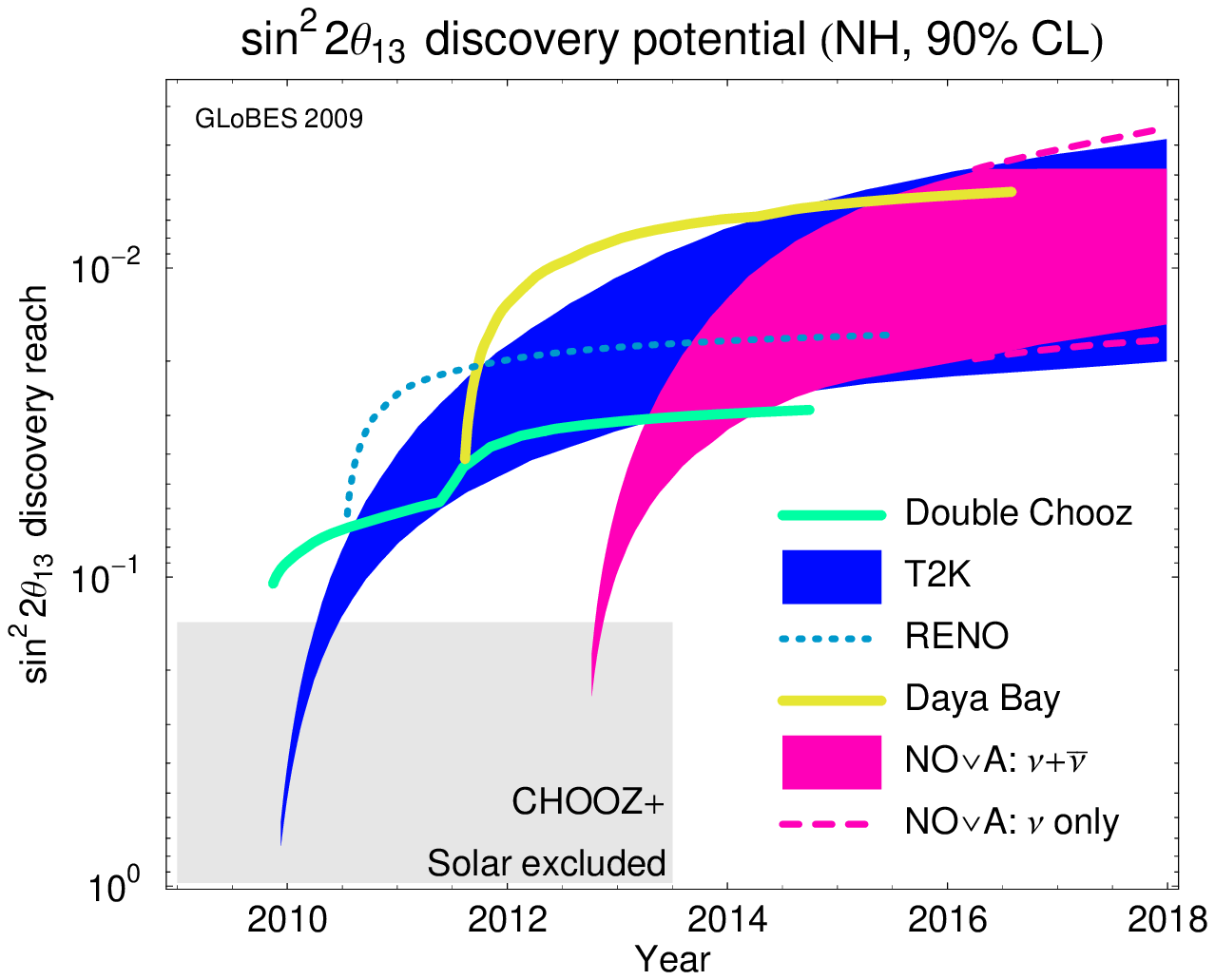} \hspace*{0.01\textwidth}
\includegraphics[width=0.48\textwidth]{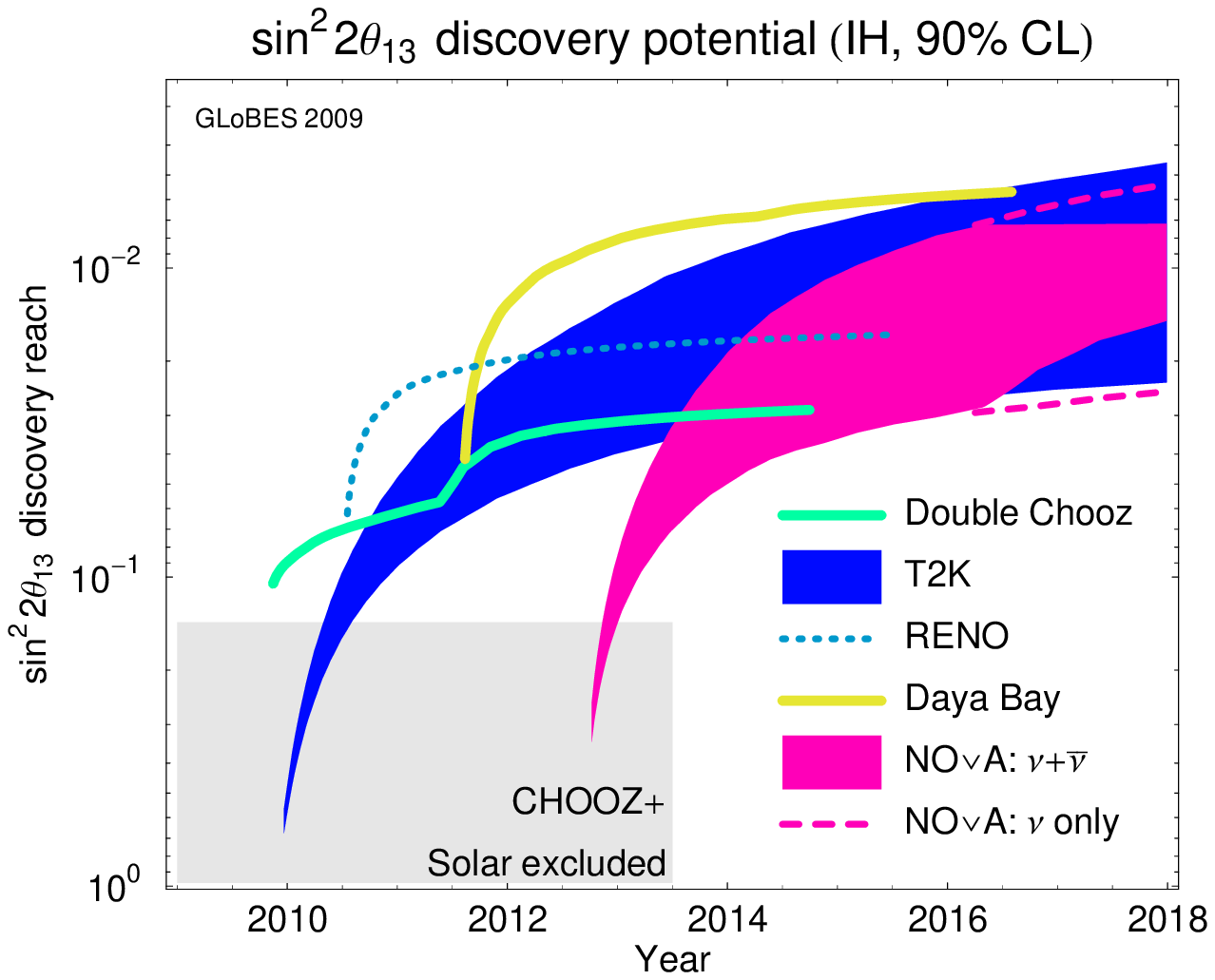}
\end{center}
\mycaption{\label{fig:evoldisc} Evolution of the $\theta_{13}$ discovery
potential as a function of time (90\% CL), \ie, the smallest value of
$\theta_{13}$ which can be distinguished from zero at 90\%~CL. We assume the
normal and inverted simulated hierarchies in the left and right panels,
respectively.  The bands reflect the (unknown) true value of $\deltacp$.}
\end{figure}

The $\theta_{13}$ discovery potential is shown in \figu{evoldisc} as a
function of time. For the beam experiments, the dependence on the true value
of $\deltacp$ is shown as shaded region, whereas the reactor experiments are
not affected by the true $\deltacp$. There is a small dependence on the true
mass hierarchy for the beam experiments, compare left and right panels.
The comparison of \Figs~\ref{fig:evoldisc} and \ref{fig:evolsens} shows that
suitable values of $\deltacp$ may significantly improve the discovery
potential of beams compared to their sensitivity limit. Indeed, the beam
experiments may discover $\theta_{13}$ for smaller $\theta_{13}$ than Daya
Bay in a small fraction of the parameter space (see also \figu{disc}).
Overall, it may however be more likely that the reactor experiments are
faster. The NO$\nu$A bands become more narrow for some additional
antineutrino running, which means that the best case potential gets slightly
worse, but the worst case becomes somewhat better. Again, this may not be
an argument for antineutrino running since the anticipated Daya Bay
sensitivity may not be exceed-able. For a more detailed discussion of the
potential antineutrino running, we refer to \Sec~\ref{sec:upgrades}.

Note that this discussion is based on the unitarity standard three-flavor
oscillation framework. If the search for new physics is taken into account,
different reactor experiments, or reactor experiments and superbeams, may
imply different information and therefore be very complementary; see, \eg,
\Refs~\cite{Schwetz:2005fy, Kopp:2007ne}. 

%%%%%%%%%%%%%%%%%%%%%%%%%%%%%%%%%%%%%%%%%%%%%%%%%%%%%%%%%%%%%%%%%%%%%%%%%%%%%
\subsection{Mass hierarchy and CP violation discovery potentials}

\begin{figure}[tp]
\begin{center}
\includegraphics[width=\textwidth]{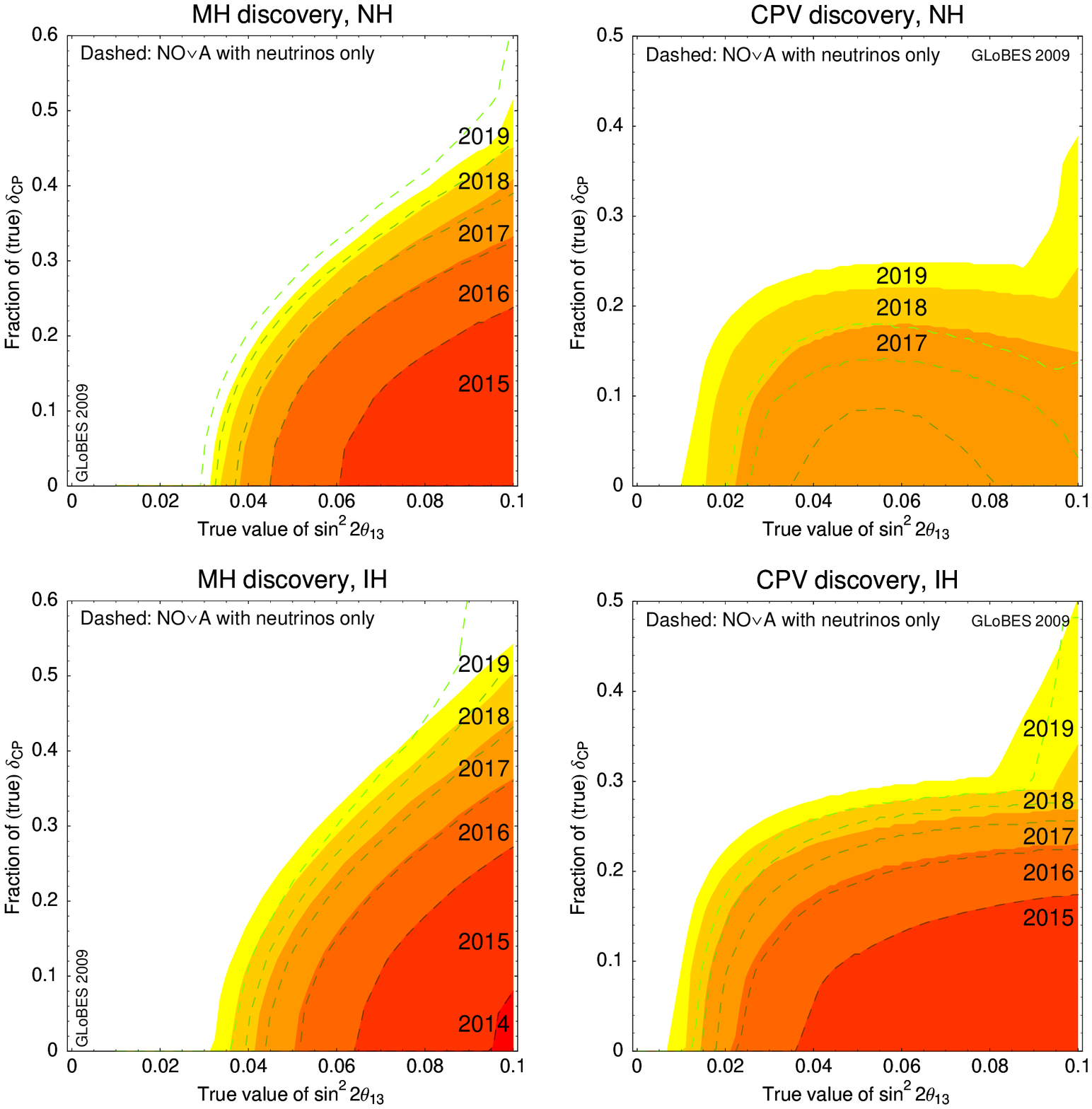}
\end{center}
\mycaption{\label{fig:tsl} Mass hierarchy (left panels) and CP violation
  (right panels) discovery potentials as a function of true $\stheta$ and
  fraction of true $\deltacp$ at 90\%~CL from T2K, NO$\nu$A and reactor
  data. The upper row corresponds to the normal simulated hierarchy, the
  lower row to the inverted simulated hierarchy. The different shadings
  correspond to different points of time, as marked in the plots (note that
  ``2015'' here means mid 2015). The dashed curves refer to NO$\nu$A with
  neutrino running only, whereas the shaded contours refer to the nominal
  NO$\nu$A neutrino-antineutrino plan. If no contour is shown for a specific
  year, there is no sensitivity. Note the different scales on the vertical
  axes.}
\end{figure}

We show in \figu{tsl} the mass hierarchy (left panels) and CP violation
(right panels) discovery potentials as a function of true $\stheta$ and
fraction of (true) $\deltacp$ from T2K, NO$\nu$A, and Daya Bay. The upper
row corresponds to the normal simulated hierarchy, the lower row to the
inverted simulated hierarchy. The different contours represent different
points in time, and can be viewed as timeslices.  Obviously, there will be
no mass hierarchy discovery before 2014, and no CP violation discovery
before 2015. In the most optimistic case without upgrades, the mass
hierarchy and CP violation can be discovered in about 50\% of all cases of
$\deltacp$ by 2019. Note, however, that we show the 90\% CL potentials here,
whereas there is only extremely poor $3\sigma$ sensitivity. Therefore, as a
first conclusion, there will be no high significance determination of the
mass hierarchy or CP violation without the next generation of experiments.
In the most optimistic case, some hints may be obtained in about 2015-2018.
As discussed earlier (\cf, \figu{disc}), NO$\nu$A plays a key role in these
discovery potentials.

As far as the NO$\nu$A switching to antineutrinos is concerned, we show in
\figu{tsl} the case if NO$\nu$A runs with neutrinos only as dashed
curves. As expectable, for CP violation the antineutrino run is mandatory.
For the mass hierarchy discovery, however, it is of secondary importance.
For a more detailed discussion, see \Sec~\ref{sec:upgrades}.

Interesting observations can be obtained from the comparison between
\figu{tsl} and the $\theta_{13}$ time evolution. If $\stheta \lesssim 0.02$,
it may already be excluded by Daya Bay early 2012 (\cf, \figu{evolsens}). In
this case, it will be clear already before the NO$\nu$A startup that there
is no chance to find the mass hierarchy or CP violation without significant
upgrades or new experiments. On the other hand, if $\stheta \gtrsim 0.02$,
this will be known at about the same time (\cf, \figu{evoldisc}), meaning
that NO$\nu$A has a realistic chance to see something. In this way, the
``branching point'' 2012 will be an interesting point of time at which
strategic decisions on the future neutrino oscillation program can be made,
such as in favor of upgrades or a new high intensity facility.

%%%%%%%%%%%%%%%%%%%%%%%%%%%%%%%%%%%%%%%%%%%%%%%%%%%%%%%%%%%%%%%%%%%%%%%%%%%%%%%%%%%%%%%
\section{Beam upgrades for large $\boldsymbol{\theta_{13}}$ and the $\boldsymbol{\nu}$-$\boldsymbol{\bar
  \nu}$ optimization}
%%%%%%%%%%%%%%%%%%%%%%%%%%%%%%%%%%%%%%%%%%%%%%%%%%%%%%%%%%%%%%%%%%%%%%%%%%%%%%%%%%%%%%%
\label{sec:upgrades}

In the previous sections we have seen that the sensitivity to CPV and MH of
the discussed experiments in their nominal configuration as defined in
Sec.~\ref{sec:experiments} is marginal, at best. In this section we address
the following question. Imagine that a finite value of $\theta_{13}$ is
discovered soon; can ``modest upgrades'' of the experiments considerably
improve the sensitivity to CPV and MH? With ``modest upgrades'' we mean
modifications of existing equipment and infrastructure. This includes a
longer running time and an upgraded beam power for both accelerator
experiments and the addition of antineutrino running in T2K.\footnote{The
anti-neutrino running beam fluxes for the 2.5 degree off-axis beam for T2K
are taken from~\cite{NAKAYA}.} It does not include new beam lines or new
detectors. In particular, our toy scenario is the following:\footnote{In
this section, if not stated otherwise, years denote always the middle of
year, \ie, 2019 = July 1st, 2019.}
\begin{description}
\item[T2K] We assume that a proton driver is installed, which
  increases the beam power from 0.75 to 1.66~MW, linearly from 2015
  to 2016~\cite{T2KUPGRADE}.
\item[NO$\nu$A] At Fermilab, the proton driver ``ProjectX'' is
  discussed. We assume a linear increase from 0.7 to 2.3~MW from March
  2018 to March 2019~\cite{PROJECTX}.
\end{description}
For Double Chooz, in principle, there exists the upgrade option to
Triple Chooz~\cite{Huber:2006vr}. However, from \figu{evolsens}, we do
not see how that could compete with Daya Bay (either timescale-wise,
or physics-wise). Therefore, we do not discuss this possibility here.
Likewise, neither for RENO nor for Daya Bay upgrade options are
discussed/anticipated.

To be specific, we assume in the following that the true value of
$\theta_{13}$ is $\stheta = 0.1$ and the true hierarchy is normal.
According to \figu{evoldisc} this value of $\theta_{13}$ will be
discovered before 2011 by Double Chooz, T2K and RENO. We focus first
on CPV and comment on the MH in Sec.~\ref{sec:upgr-evol}.

%%%%%%%%%%%%%%%%%%%%%%%%%%%%%%%%%%%%%%%%%%%%%%%%%%%%%%%%%%%%%%%%%
\subsection{Optimization of neutrino/antineutrino running}

\begin{figure}[tp]
\begin{center}
\includegraphics[width=\textwidth]{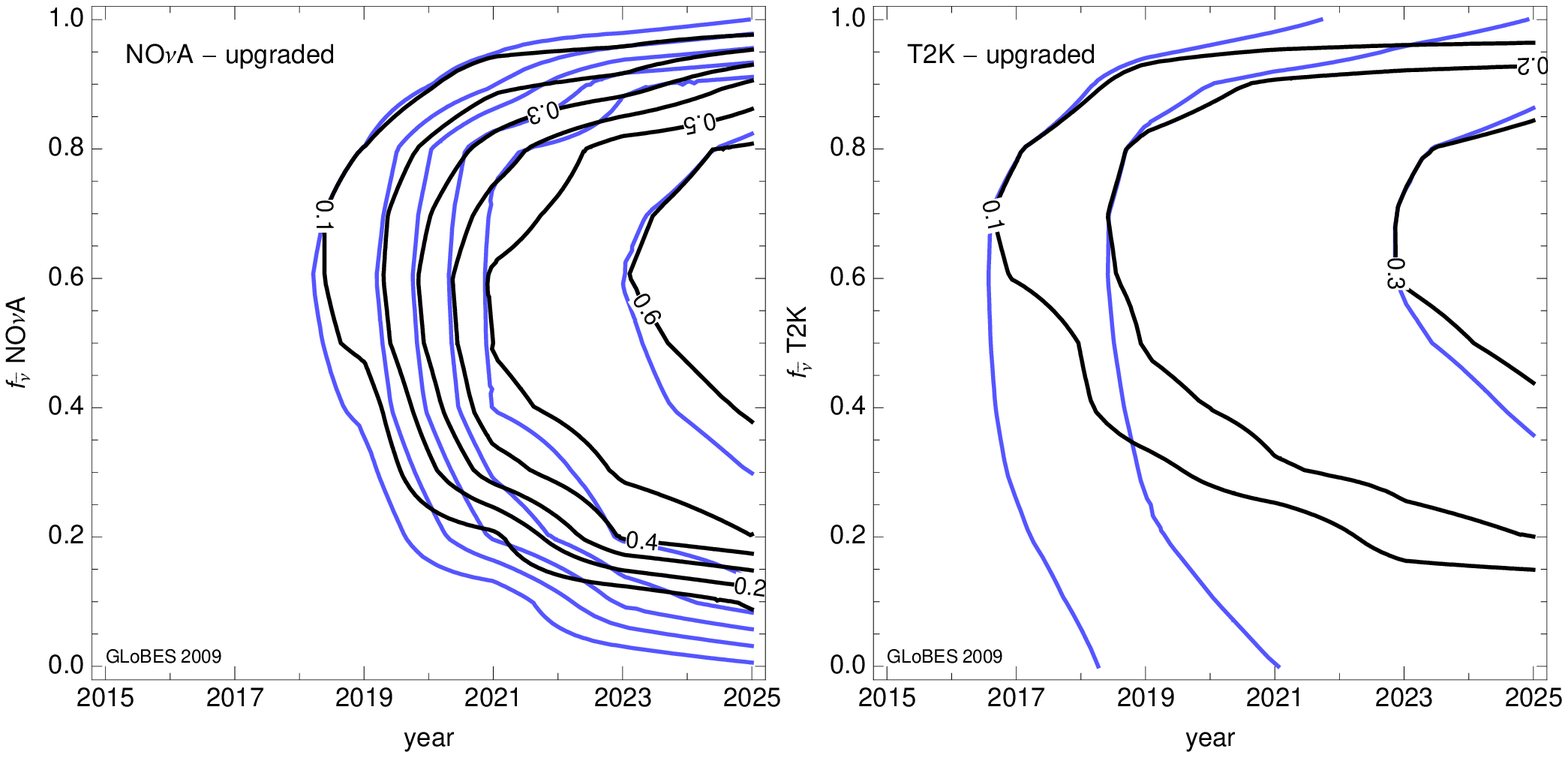}
\end{center}
\mycaption{\label{fig:ns} The contours show the CP fraction for which
  CP violation can be established at 90\%~CL as a function of the year
  and the fraction of antineutrino running $f_{\bar\nu}$ up to that
  year. We assume the upgraded run plan. The black contours assume T2K
  (right hand panel) or NO$\nu$A (left hand panel) data only, whereas
  the colored contours show the result for the case where reactor data
  on $\theta_{13}$ is included.}
\end{figure}

We start our analysis by discussing the optimization of the fraction of
neutrino/antineutrino exposures. \figu{ns} shows the fraction of
$\deltacp$-values for which CPV can be established at 90\%~CL as a function
of time and as a function of the fraction of antineutrino running
$f_{\bar\nu}$ for NO$\nu$A (left) and T2K (right). Here, $f_{\bar\nu} = 0 \,
(1)$ corresponds to the full exposure with neutrinos (antineutrinos). As
expected, in order to address CPV a sizable fraction of antineutrino running
is required: $f_{\bar\nu}=0.6-0.7$ for T2K and $f_{\bar\nu}=0.5-0.6$ for
NO$\nu$A. The plot is for normal hierarchy, but results are similar for
inverted hierarchy. Without reactor data $f_{\bar\nu}=0$ is strongly
disfavored even when the beams are upgraded, indicating that spectral
information is not enough to provide a measurement of $\deltacp$. 

NO$\nu$A with upgrades reaches a CP fraction of about 60\%, while T2K
achieves only about 30\%. The reason is that the beam upgrade is a
factor of 3.2 for NO$\nu$A whereas it is only 2.2 for T2K. Moreover,
antineutrino running is very difficult for T2K due to the low beam
energy. Since the ratio of antineutrino/neutrino cross sections as
well as fluxes decreases with energy, statistics is rather poor for
T2K antineutrino data. This is especially true for the $2.5^\circ$
off-axis T2K beam. Optimization of beam optics or energy is beyond the
scope of this work.

The blue (light gray) curves in \figu{ns} show the case when reactor data on
$\theta_{13}$ are combined with that of the beam experiment. While for NO$\nu$A,
reactor data have only a minor impact, for T2K, the combination
with a reactor experiment can considerably relax the requirement on
$f_{\bar\nu}$ and even replace the antineutrino running  altogether in most
cases~\cite{Minakata:2003wq}. However, the optimal CPV fraction for
T2K is not improved significantly by adding reactor data.

\begin{figure}[tp]
\begin{center}
\includegraphics[width=\textwidth]{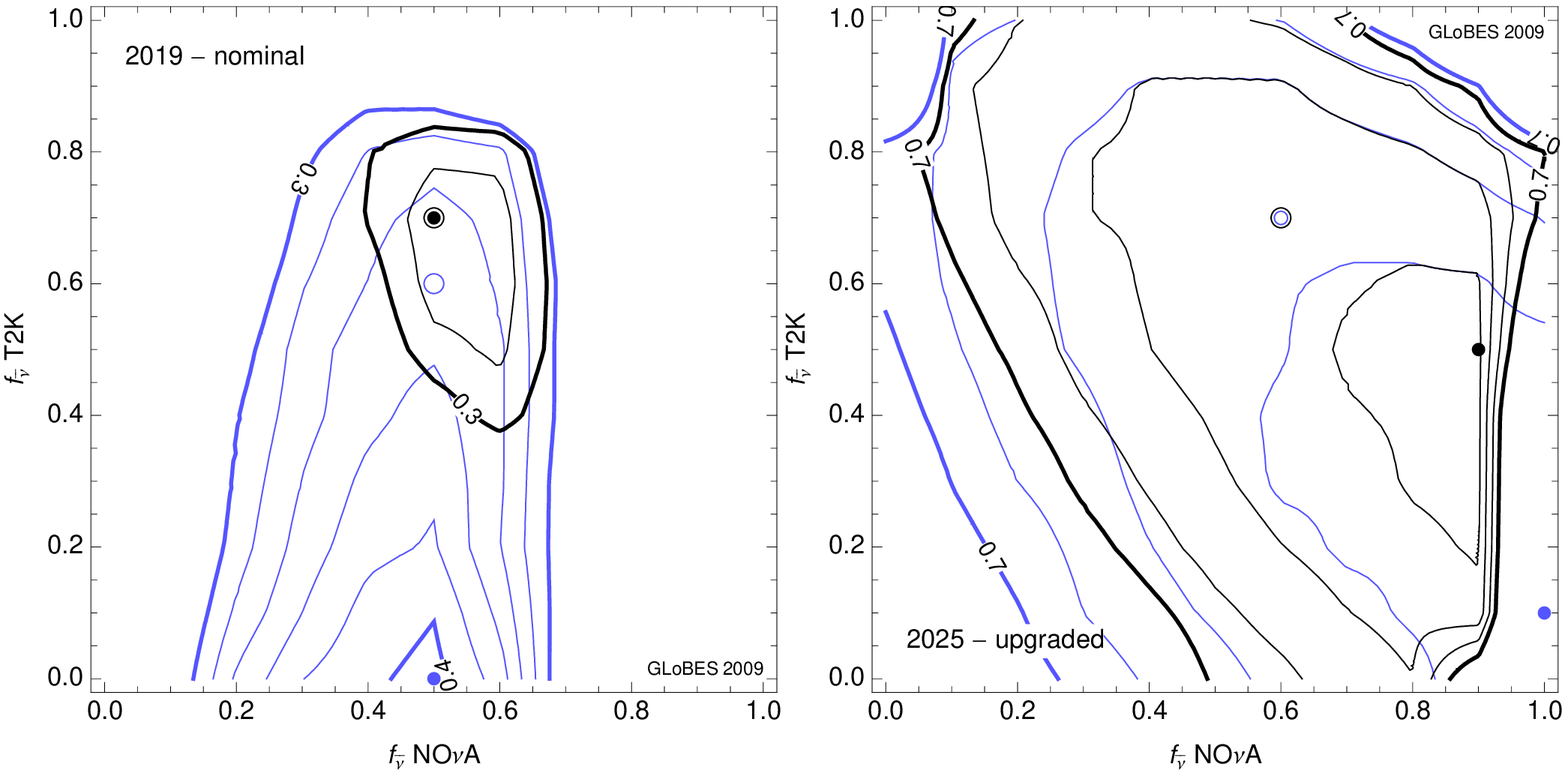}
\end{center}
\mycaption{\label{fig:anuT2KvsN} The contours show the CP fraction for
  which CP violation can be established at 90\%~CL as a function of
  the antineutrino fraction in NO$\nu$A and T2K in the year 2019
  assuming the nominal run plan (left hand panel) or in the year 2025
  with the upgraded run plan (right hand panel). The black contours
  assume the combined data from NO$\nu$A and T2K only, whereas for the
  colored contours also reactor data on $\theta_{13}$ is included.
  The thick contours have a spacing of $0.1$ whereas the thin
  contours have a spacing of $0.02$. The black disk denotes the
  optimal point for beam data only and whereas the colored disk
  denotes the optimal point for beam and reactor data combined. The
  circles denote the optima found for each experiment individually.}
\end{figure}

\figu{anuT2KvsN} shows the neutrino/antineutrino optimization for the
combined data of T2K and NO$\nu$A, with and without reactor data.  In the
left panel we assume only the nominal luminosities for the beams, but allow
for an arbitrary fraction of antineutrinos for both experiments. Without
reactor data, the optimum is at $f_{\bar\nu}^\mathrm{NO\nu
A}/f_{\bar\nu}^\mathrm{T2K} = 0.5/0.7$. The CP fraction is about 32\% and
antineutrino data is required for both experiments. If reactor data are
added, the optimum is at $f_{\bar\nu}^\mathrm{NO\nu
A}/f_{\bar\nu}^\mathrm{T2K} = 0.5/0$, which is remarkably close to the
nominal plan for the experiments. The CP fraction is 40\%, in agreement with
\figu{tsl}. We observe again that reactor data are more useful
than the statistically weak antineutrino data in T2K, which means that it is preferable
to fully explore neutrino data in T2K.

The right panel of \figu{anuT2KvsN} shows the same analysis for the upgraded
beams. The tendency is to use large antineutrino fractions in NO$\nu$A and
less in T2K, the optimum being $f_{\bar\nu}^\mathrm{NO\nu
A}/f_{\bar\nu}^\mathrm{T2K} = 0.9/0.5 \,(1/0.1)$ without (with) reactors.
We find an optimal CP sensitivity at 90\%~CL for beam+reactor data for about
76\% of all $\deltacp$ values. The landscape becomes rather flat and similar
CP fractions can be obtained for a wide range of antineutrino fractions.

%%%%%%%%%%%%%%%%%%%%%%%%%%%%%%%%%%%%%%%%%%%%%%%%%%%%%%%%%%%%%%%%%%%%%%%%%%%
\subsection{Optimal run plan}

\begin{figure}[tp]
\begin{center}
\includegraphics[width=\textwidth]{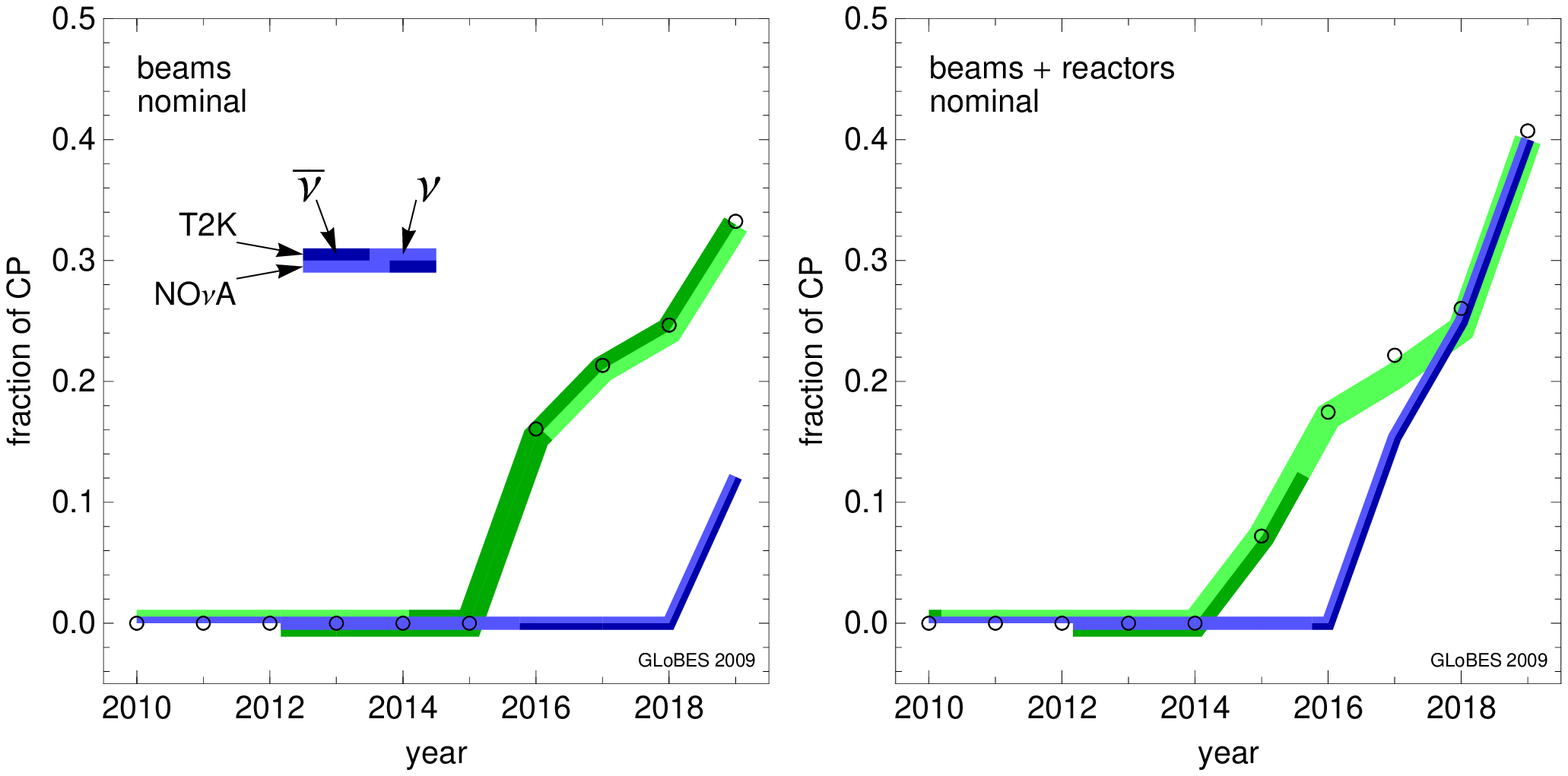}
\includegraphics[width=\textwidth]{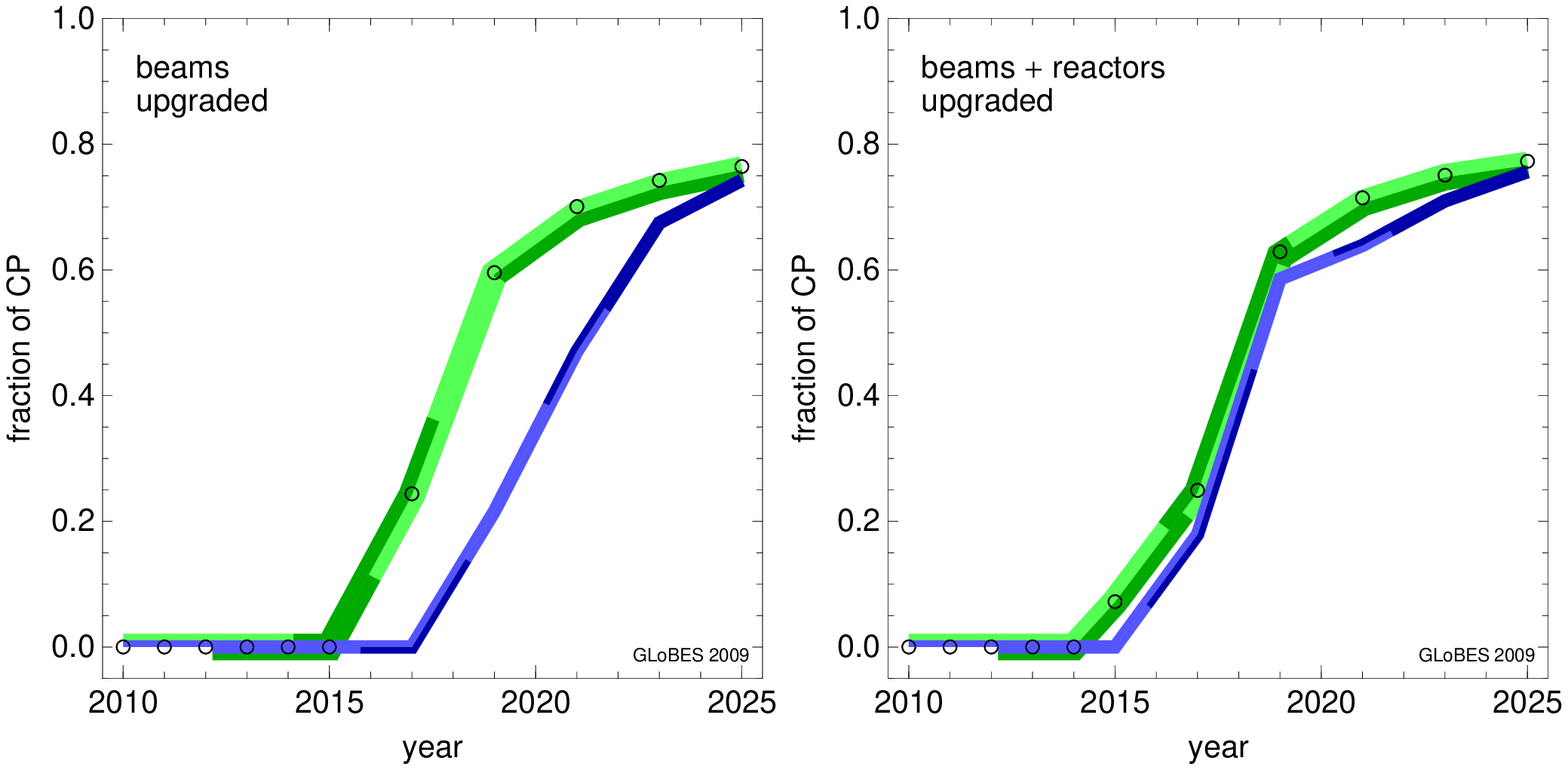}
\end{center}
  \mycaption{\label{fig:nopt} Fraction of $\deltacp$ values for which CP
  violation can be discovered at 90\%~CL as a function of time for the
  nominal run plan (upper row) and the upgraded run plan (lower row). These
  plots assume the true sign of $\Delta m_{31}^2>0$. The left hand column
  shows the results for a combination of T2K and NO$\nu$A alone, whereas the
  right hand column includes reactor data.  The green band is for the
  globally optimal run plan, whereas the blue band is for the nominal run
  plan. Light and dark colors indicate neutrino and antineutrino running,
  respectively. The black circles denote the absolute maximal performance
  for a given time.}
\end{figure}

In the previous subsection we investigated the optimal total fraction of
antineutrino running for the full lifetime of the experiment. In this
section, we proceed further and ask the following question: what is
the neutrino/antineutrino running strategy such that at {\it at each
point in time} the optimal sensitivity to CPV is obtained? 
Let us first define a reference neutrino-antineutrino scenario (``nominal
run plan'') as follows:
\begin{description}
\item[Phase~I: Until design luminosity reached] T2K: neutrinos only,
  NO$\nu$A: 50\% neutrinos followed by 50\% antineutrinos (such as in
  LOIs)
\item[Phase~II: After design luminosity reached until mid 2025] T2K:
  50\% neutrinos followed by 50\% antineutrinos. NO$\nu$A: 50\%
  neutrinos followed by 50\% antineutrinos
\end{description}
Note that the definition of phase~I and~II is not connected with the upgrade
point of time, but the design luminosity. The $\deltacp$ fractions for CPV
at 90\%~CL for this nominal run plan are shown in \figu{nopt} as a function
of time with blue bands, where light and dark colors indicate neutrino and
antineutrino running, respectively.

Now we investigate whether a globally optimized neutrino-antineutrino
run plan has the potential to improve the performance with respect to
this reference plan. This (non-deterministic) optimization is
performed with the help of a genetic algorithm described in detail in
appendix~\ref{app:genetics}. Roughly, the algorithm evolves a set of
switching times between neutrinos and antineutrinos for T2K and
NO$\nu$A by favoring ``individuals'' with the largest CP fraction at
each point in time and disfavoring the ones with too many switching
points. After evolving a randomly chosen population of 2000
individuals for 50 generations we average over the 100 ``fittest''
individuals.  The results of this optimization search are shown as
green bands in \figu{nopt}, see also Tab.~\ref{tab:optimal} in the
appendix.

Let us first discuss the nominal exposures without the beam upgrades
(upper row of panels). We find that the reference $\nu$-$\bar\nu$ run
plan outlined above provides a very suboptimal performance and can
delay physics by four to five years for beam data without reactors. If
reactor data are added, delays will be reduced to one or two years.
Again we observe that reactor data allow to avoid antineutrino running
in T2K, and we find that an early antineutrino run in NO$\nu$A is
preferred. The optimal strategy is not costly in terms of switches,
only one or two additional switches are required.

The lower panels in \figu{nopt} show the results for upgraded beams.
Again the reference $\nu$-$\bar\nu$ run plan provides a suboptimal
performance and can delay physics by two to three years for beam data
only. Inclusion of reactor data helps a lot and delays are reduced to
one year. Here reactor data no longer allow for circumventing
antineutrino running in T2K. We observe that the final overall
performance with and without reactors is very similar if the run plan
is optimized. Thus, it seems that one should strive to optimize the
beams as much as possible on their own. In that way the reactor data
can be used as an independent cross check of the result, for instance,
to provide some
sensitivity to non-standard neutrino physics. Again we find that the
optimal strategy prefers early antineutrino data from NO$\nu$A, and
that only one or two additional switches are required.

%%%%%%%%%%%%%%%%%%%%%%%%%%%%%%%%%%%%%%%%%%%%%%%%%%%%%%%%%%%%%%%%%%%%%%%
\subsection{Time evolution of physics potential}
\label{sec:upgr-evol}

\begin{figure}[tp]
\begin{center}
\includegraphics[width=\textwidth]{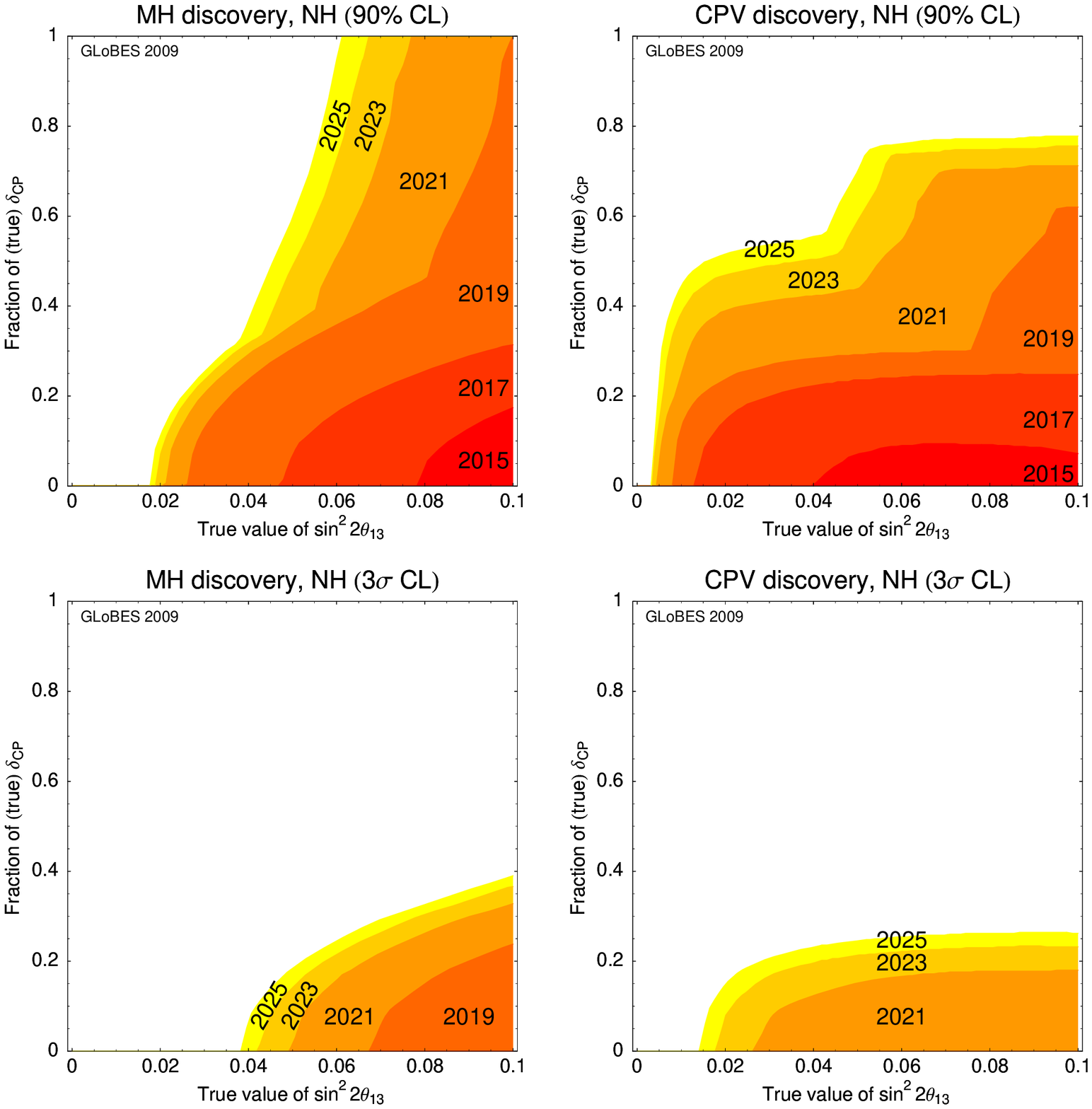}
\end{center}
\mycaption{\label{fig:tslUPGR} Mass hierarchy (left panels) and CP violation
  (right panels) discovery potentials as a function of true $\stheta$ and
  fraction of (true) $\deltacp$ for our optimal run plan including upgrades
  and reactor experiments; \cf, \Tab~\ref{tab:optimal}. The upper panels are
  for 90\% CL, the lower panels for $3 \sigma$ CL.  The different shadings
  corresponds to different points of time, as marked in the plots. Note the
  different scales on the vertical axes compared to \figu{tsl}.}
\end{figure}

Having optimized the CPV performance for a true value $\stheta = 0.1$
in the previous subsections, we now relax this
assumption\footnote{Note, that in general the $\chi^2$-functions will
  have lower values and hence be less steep for smaller $\theta_{13}$.
  Therefore, optimization is less crucial and a wider range of running
  fractions is acceptable. Thus a solution which was optimal for large
  $\theta_{13}$ will still be very close to optimal for somewhat
  smaller $\theta_{13}$. However, all sensitivities will be
  decreased.} and discuss the sensitivities of these optimized
configurations as a function of $\theta_{13}$. We consider the
upgraded beams for T2K and NO$\nu$A combined with reactor data, and
show results for CPV as well as for the neutrino mass hierarchy.
\figu{tslUPGR} shows the discovery potential as a function of true
$\stheta$ and fraction of true $\deltacp$ for times from 2015 to 2025.
The upper row of this figure shows the discovery potentials at the
90\%~CL.  These results can be compared to \figu{tsl} (upper panels),
where one should note the different scales on the vertical axes.
Obviously, with the optimal upgrade plan, there is a significant
improvement of the MH and CPV discovery potentials. At the 90\%
confidence level, there will be hints for the MH and CPV for $\stheta
\gtrsim 0.05$ for most values of $\deltacp$ around 2025.

However, certainly a 90\%~CL is not sufficient to make any meaningful
statement about a discovery. Therefore, we show in the lower row of
\figu{tslUPGR} the corresponding results at $3\sigma$~CL. Obviously the
sensitivity regions reduce drastically, however, we see from the figure that
assuming both beams upgraded, a fully optimized neutrino/antineutrino run
plan, and data from reactors a non-negligible discovery potential at
$3\sigma$ will be reached in 2025. The mass hierarchy can be identified for
$\stheta \gtrsim 0.05$ for about 20\% to 40\% of $\deltacp$ values, whereas
CPV can be discovered for $\stheta \gtrsim 0.02$ for 25\% of $\deltacp$
values. In both cases, MH and CPV, there is sensitivity for values of $\deltacp$ 
around $3\pi/2$ ($\pi/2$) if the true hierarchy is normal (inverted). This
behaviour is visible in Figs.~\ref{fig:deltatheta} and \ref{fig:deltathetai}
and it is related to the sign of the matter effect, see, \eg,
\Ref~\cite{Winter:2003ye} for a discussion.

Let us mention that in the previous two subsections we have optimized the
sensitivity to CPV. However, it turns out that this is also very close to
the optimal performance for the MH determination. Furthermore, it is a
feature of our optimization plan that the performance in the case of a
(true) inverted hierarchy is similar to one for the normal hierarchy.
Therefore it is not shown explicitely here.

%%%%%%%%%%%%%%%%%%%%%%%%%%%%%%%%%%%%%%%%%%%%%%%%%%%%%%%%%%%%%%%%%%%%%%%%%%%%%%%%%%%%%%%%%
\section{Summary and conclusions}
%%%%%%%%%%%%%%%%%%%%%%%%%%%%%%%%%%%%%%%%%%%%%%%%%%%%%%%%%%%%%%%%%%%%%%%%%%%%%%%%%%%%%%%%%
\label{sec:conclusions}

In this work we have discussed the physics potential of the upcoming
neutrino oscillation experiments Double Chooz, Daya Bay, RENO, T2K,
and NO$\nu$A. In the first part we have reconsidered the sensitivities
from their nominal exposures, such as stated in the experimental
proposals. These results are summarized in \Tab~\ref{tab:summary}.
While Double Chooz and maybe RENO could allow for a fast discovery of
a relatively large value of $\theta_{13}$, the ultimate reactor
experiment will be Daya Bay, which will dominate a few months after it
comes online. The $\theta_{13}$ performance of the beam experiments
will depend on the value of $\deltacp$ and the performance indicator.
In the case of no signal, the limit on $\theta_{13}$ will be similar
to the one from Double Chooz, whereas a favorable value of $\deltacp$
will allow for a discovery for slightly smaller $\theta_{13}$ than
Daya Bay.

\begin{table}
\centering
\begin{tabular}{lccc}
\hline
  & $\stheta$ & $|\ldm|$ & $|\sin^2\theta_{23} - 0.5|$ \\
\hline
  Double Chooz & 0.033 (0.060) & -- & --\\
% MORE PRECISELY: 0.0326 (0.0596)
  T2K & 0.004--0.027 (0.011--0.040) & $^{+2.0\%}_{-1.9\%}$ ($^{+3.7\%}_{-3.5\%}$) & 0.055 (0.074)\\
  RENO & 0.018 (0.033) & -- & --\\
% MORE PRECISELY: 0.0179 (0.0328)
  Daya Bay  & 0.007 (0.012) & -- & --\\
% MORE PRECISELY: 0.0066 (0.0120)
  NO$\nu$A & 0.005--0.014 (0.011--0.025) & $^{+2.5\%}_{-2.0\%}$ ($^{+4.7\%}_{-3.6\%}$) & 0.065 (0.092) \\
\hline
\end{tabular}
  \mycaption{Summary of nominal sensitivities at 90\% ($3\sigma$). We show
  the discovery potential for $\stheta$ (where for the beams we give the
  best and worst sensitivity depending on $\deltacp$ and mass hierarchy),
  the accuracy on $|\ldm|$ (for maximal mixing), and the sensitivity to
  deviations from maximal $\theta_{23}$ mixing. \label{tab:summary}}
\end{table}

We have found that the global sensitivity to CP violation and the
neutrino mass hierarchy of all experiments with nominal exposures is
marginal. For the largest allowed values of $\theta_{13}$, typically a
hint at 90\%~CL can be obtained for about 25\% to 50\% of all possible
values of $\deltacp$, while almost nothing can be said at $3\sigma$.
Therefore, we have investigated the possibilities to increase the
sensitivity by minor upgrades of the beam experiments in case
$\theta_{13}$ is not too far from its present bound and hence
discovered soon. These upgrades are based on existing equipment and
include an increase of beam power with the help of proton drivers,
longer running times, and the addition of antineutrino running in T2K.
We have performed an optimization study concerning the distribution of
neutrino and antineutrino data runs in T2K and NO$\nu$A in order to
maximize the global sensitivity reach. We have found that typically
communication between the two beam experiments will improve the
overall sensitivity. For example, the optimal sensitivity usually
requires relatively early antineutrino data from NO$\nu$A.
Furthermore, we have found that the sensitivity of the optimized
running strategy for T2K plus NO$\nu$A is rather similar to the one
with reactor experiment data used in addition.

These results are summarized in \figu{tslUPGR}. Assuming both beams
upgraded, a fully optimized neutrino/antineutrino run plan, and data from
reactors, we have found that a hint for CP violation at 90\%~CL can be
obtained around 2025 for a large fraction of $\deltacp$ for a reasonably
large $\theta_{13}$.  However, the discovery potential for CP violation and
mass hierarchy drastically reduces at the $3\sigma$~CL. Nevertheless, a
non-negligible discovery potential at $3\sigma$ will be reached in 2025: The
mass hierarchy can be identified for $\stheta \gtrsim 0.05$ for about 20\%
to 40\% of $\deltacp$ values, whereas CP violation can be discovered for
$\stheta \gtrsim 0.02$ for 25\% of $\deltacp$ values. Sensitivity to the
mass hierarchy and CP violation is obtained for values of $\deltacp$ close
to maximal CP violation: $\deltacp \simeq 3\pi/2 \, (\pi/2)$ for a true
normal (inverted) hierarchy, but not for the opposite case.

Let us mention that NO$\nu$A dominates the MH determination and typically
also has a somewhat better sensitivity than T2K for CPV. For example, at
$\stheta=0.1$, NO$\nu$A alone obtains a CP fraction for CPV at 90\%~CL of
about 60\%, whereas T2K reaches 35\%, and the combination yields 76\% (both
beams with upgrades). The reasons are that the beam upgrade we consider is a
factor of 3.2 for NO$\nu$A whereas it is only 2.2 for T2K, and antineutrino
running is very difficult for T2K due to the low beam energy. Since the
ratios of antineutrino/neutrino cross sections as well as fluxes decrease
with energy, statistics is rather poor for T2K antineutrino data.

Our results raise the question on how to adapt the global oscillation
strategy in case $\theta_{13}$ were discovered soon. Although ``minor
upgrades'' of existing facilities may provide a non-negligible sensitivity
to CP violation and the mass hierarchy, there is high risk associated with
this strategy, since for about 75\% of all possible values for $\deltacp$ no
discovery will be possible at the 3$\sigma$ level. 
In contrast, a high precision facility such as a wide band superbeam or a
beta beam with very large detectors or a neutrino factory~\cite{euronu, ids}
would certainly be able to perform a solid determination of CP violation and
the mass hierarchy in case of such a large value of $\theta_{13}$. 
Therefore, we conclude that the upcoming generation of oscillation
experiments may lead to interesting indications for the mass hierarchy and
CP violation, but it is very likely that an experiment beyond the upcoming
superbeams (including reasonable upgrades) will be required to confirm these
hints.

\subsubsection*{Acknowledgments}

This work was supported by the European Union under the European Commission
Framework Programme~07 Design Study EUROnu, Project 212372, and by the
Transregio Sonderforschungsbereich TR27 ``Neutrinos and Beyond'' der
Deutschen Forschungsgemeinschaft. WW also would like to acknowledge support
from the Emmy Noether program of Deutsche Forschungsgemeinschaft, contract
WI 2639/2-1.

%%%%%%%%%%%%%%%%%%%%%%%%%%%%%%%%%%%%%%%%%%%%%%%%%%%%%%%%%%%%%%%%%%%%%%%%%%%%%%%
\begin{appendix}
%%%%%%%%%%%%%%%%%%%%%%%%%%%%%%%%%%%%%%%%%%%%%%%%%%%%%%%%%%%%%%%%%%%%%%%%%%%%%%%
\section{Optimization method}
\label{app:genetics}

Luminosity $\mathcal{L}$ is given by the time integrated product of target
mass $m$ and beam power $p$
\begin{equation}
\mathcal{L}(t):=\int_{t_0}^t\, dt'\,m(t')\,p(t')\,,
\end{equation}
where we allow for $m$ and $p$ to be time dependent. We define the
normalized luminosity $L$ by
\begin{equation}
L(t)=\frac{\mathcal{L}(t)}{\mathcal{L}_0}\,,
\end{equation}
where $\mathcal{L}_0$ corresponds to the design luminosity of the
experiment. Since, in the following we will discuss T2K and NO$\nu$A
every quantity pertaining to T2K will carry $T$ as subscript and those
carrying subscript $N$ pertain to NO$\nu$A.

T2K and NO$\nu$A both can run neutrinos or antineutrinos. We will
denote the the fraction of antineutrinos in each experiment by
$f$, where
\begin{equation}
f(t)=\frac{L_{\bar\nu}(t)}{L(t)}\,.
\end{equation}
The current run plans foresee no antineutrino running for T2K,
$f_{T}^0=0$ and 50\% of antineutrino running for NO$\nu$A,
$f_{N}^0=0.5$. For a given, true value of $\sin^22\theta_{13}$ we can
compute the obtainable fraction of $\deltacp$, $\mathcal{C}$, for the
measurement of CP violation as a functions of both $f_{T}$ and
$f_{N}$. This is shown in figure~\ref{fig:anuT2KvsN}. Of course this
will be a time dependent quantity denoted by
$\mathcal{C}(f_N(t),f_T(t))$. To simplify our notation we will write
$\mathcal{C}_t(f_N,f_T)$. For algorithmic purposes, it turns out to be
useful to define the normalized CP fraction $C$ by
\begin{equation}
C_t(f_N,f_T)=\frac{\mathcal{C}_t(f_N,f_T)}{\hat{\mathcal{C}}_t}\quad\text{with}\quad\hat{\mathcal{C}}_t=\max_{f_N,f_T}
\mathcal{C}_t(f_N,f_T)\,.
\end{equation}
Thus, $C_t$ will have values from $0$ to $1$ for all times $t$. Note,
that there will be seperate and distinct $\mathcal{C}_t$ and $C_t$ for
the true hierarchy normal (NH) or inverted (IH).

Since an experiment needs to reverse the polarity of the electric
current in the focusing horn, its operation will be divided into
phases of neutrino running and those of antineutrino running. Any
such division can be described a set of times at which the operation
mode switches, plus the initial running configuration. We will denote
such a set as $g=\{g_1,\ldots,g_{l_g}\}$ for initial neutrino running and
with $\bar{g}=\{g_1,\ldots,g_{l_{\bar g}}\}$ for initial antineutrino
running. The $g_i$ denote the individual switching times and $l_g$ the
total number of switches. $g$ uniquely
determines the the antineutrino fraction as a function of time

\begin{equation}
f_g(t)=\frac{L(t)-L(g_{k+1})}{L(t)}+\sum_{i=0}^{2i<k}\frac{L(g_{2i+1})-L(g_{2i})}{L(t)}\quad\text{with}\quad
g_{k+1}\leq t\,,
\end{equation}
for starting with antineutrinos otherwise the sum starts at
$i=1$. Thus, we now can fix the value of $C_t$ for any $t$
by inserting the $f_g(t)$ like this
\begin{equation}
\kappa(t,g_N,g_T)=C_t\left(f_{g_N}(t),f_{g_T}(t)\right)\,,
\end{equation}
in case both experiments started with neutrino running. Again, there
are two functions $\kappa^{NH}(t,g_N,g_T)$ for true hierarchy normal
and $\kappa^{IH}(t,g_N,g_T)$ for true hierachy inverted. The goal now
is to find a sequence of switching times $g$ which yields
$\kappa^{NH}(t,g_N,g_T)=1,\,\forall t$ \emph{and}
$\kappa^{IH}(t,g_N,g_T)=1,\,\forall t$. In this case, we would have
optimal sensitivity at any given moment in time, irrespective of the
true mass hierarchy. We will search such $g$ or $\bar g$ by using a
genetic algorithm. To this end, we define a fitness function to
evaluate the relative performance of different sequences $g$.

\begin{equation}
\pi(g_N,g_T)=\underbrace{\frac{1}{\max\left\{ 1-\kappa^{NH}(\tau_i,g_N,g_T),1-\kappa^{IH}(\tau_i,g_N,g_T)\right\}}}_{=:\mu^{-1}}\,\underbrace{
\frac{1}{l_{g_N}+l_{g_T}}}_{l^{-1}}\,,
\end{equation}
where the $\tau_i$ are suitably chosen points in time, {\it e.g.}
every two years from $t_0$ on. The first terms selects those solutions
which manage to balance the deviation from optimal performance at all
times. The second term penalizes solution which switch very frequently
without gaining significantly in the first term. The genotype of each
individual is given by a pair of $\{g_N,g_T\}$. In our implementation
of the algorithm the length of each sequence $l_{g_N}$ and $l_{g_T}$
are free parameters and subject to evolution. We start with a
population of widely varying sequence length. Thus, we do not have to
specify the number of switches. Also, two times $g_i$ and $g_{i+1}$ are
merged if they are less than a month apart. Each population consists
of 2000 individuals which are evolved over 50 generations. We consider
4 populations 
\begin{equation}
\{g_N,g_T\},\,\{\bar g_N,g_T\},\,\{g_N,\bar g_T\},\,\{\bar g_N,\bar g_T\}\,.
\end{equation}
The results for each population are shown in table~\ref{tab:optimal}. The
$\tau_i$ are 

$\{2010.5, 2011.5, 2012.5, 2013.5, 2014.5, 2015.5, 2016.5,2017.5,
2018.5, 2019.5\}$ 

\noindent for nominal luminosity and 

$\{2010.5, 2011.5, 2012.5, 2013.5, 2014.5, 2015.5, 2017.5,
2019.5, 2021.5, 2023.5, 2025.5\}$ 

\noindent for upgraded luminosity.

\begin{table}[t]
\begin{center}
\begin{tabular}{|c|c|c|c|c|}
\hline
\multicolumn{5}{|c|}{beams nominal}\\
  \hline
  &$g_N$/$\bar g_N$&$g_T$/$\bar g_T$&$l$&$\mu$\\
  \hline
  $\{\bar g_N,\bar g_T\}$&2016.62&2012.55, 2015.85&3&0.017\\
  $\underline{\{\bar g_N,g_T\}}$&2016.60&2014.59&2&0.017\\
  $\{g_N,\bar g_T\}$&2012.22, 2016.62&2012.41, 2015.73&4&0.017\\
  $\{ g_N,g_T\}$&2012.29, 2016.62& 2014.59&3&0.017\\
  \hline
\multicolumn{5}{|c|}{beams nominal + reactors}\\
  \hline
  &$g_N$/$\bar g_N$&$g_T$/$\bar g_T$&$l$&$\mu$\\
  \hline
  $\underline{\{\bar g_N,\bar g_T\}}$&2016.04&2010.69&2&0.067\\
  $\{\bar g_N,g_T\}$&2016.09&2017.26&2&0.058\\
  $\{g_N,\bar g_T\}$&2013.25, 2016.22&2010.77&3&0.066\\
  $\{ g_N,g_T\}$&2012.95, 2016.05& 2015.89, 2017.25&4&0.041\\
\hline
\multicolumn{5}{|c|}{beams upgraded}\\
  \hline
  &$g_N$/$\bar g_N$&$g_T$/$\bar g_T$&$l$&$\mu$\\
  \hline
  $\{\bar g_N, \bar g_T\}$&2017.07&2010.7, 2015.29&3&0.015\\
  $\underline{\{\bar g_N,g_T\}}$&2016.46, 2019.48&2014.63, 2018.17&4&0.019\\
  $\{g_N,\bar g_T\}$&2011.61, 2016.56,2019.54&2010.13, 2014.65,2018.15&6&0.019\\
  $\{ g_N,g_T\}$&2010.52, 2016.44, 2020.13& 2014.64,2017.09,2018.35&6&0.031\\
  \hline
\multicolumn{5}{|c|}{beams nominal + reactors}\\
  \hline
  &$g_N$/$\bar g_N$&$g_T$/$\bar g_T$&$l$&$\mu$\\
  \hline
  $\{\bar g_N,\bar g_T\}$&2017.09,2019.45&2010.25&3&0.032\\
  $\underline{\{\bar g_N,g_T\}}$&2017.13, 2019.44&2016.81, 2019.83&4&0.004\\
  $\{g_N,\bar g_T\}$&2011.66, 2017.23, 2018.52&2010.00, 2016.82, 2019.99&6&0.005\\
  $\{ g_N,g_T\}$&2010.88, 2017.1, 2019.46& 2016.81, 2019.83&5&0.004\\
  \hline
\end{tabular}
\end{center}
\caption{\label{tab:optimal} Average over the 100 fittest individuals in
  each population. The options shown in \figu{nopt} are underlined.}
\end{table}

\end{appendix}

%\clearpage
\bibliographystyle{apsrev}
\bibliography{references}

\begin{thebibliography}{10}
\expandafter\ifx\csname bibnamefont\endcsname\relax
  \def\bibnamefont#1{#1}\fi
\expandafter\ifx\csname bibfnamefont\endcsname\relax
  \def\bibfnamefont#1{#1}\fi
\expandafter\ifx\csname url\endcsname\relax
  \def\url#1{\texttt{#1}}\fi
\expandafter\ifx\csname urlprefix\endcsname\relax\def\urlprefix{URL }\fi
\providecommand{\bibinfo}[2]{#2}
\providecommand{\eprint}[2][]{\url{#2}}

\bibitem{Cleveland:1998nv}
\bibinfo{author}{\bibfnamefont{B.~T.} \bibnamefont{Cleveland}} \emph{et~al.},
  \bibinfo{journal}{Astrophys. J.} \textbf{\bibinfo{volume}{496}},
  \bibinfo{pages}{505} (\bibinfo{year}{1998}).

\bibitem{Altmann:2005ix}
\bibinfo{author}{\bibfnamefont{M.}~\bibnamefont{Altmann}} \emph{et~al.}
  (\bibinfo{collaboration}{GNO}), \bibinfo{journal}{Phys. Lett.}
  \textbf{\bibinfo{volume}{B616}}, \bibinfo{pages}{174} (\bibinfo{year}{2005}),
  \eprint{hep-ex/0504037}.

\bibitem{Hosaka:2005um}
\bibinfo{author}{\bibfnamefont{J.}~\bibnamefont{Hosaka}} \emph{et~al.}
  (\bibinfo{collaboration}{Super-Kamkiokande}), \bibinfo{journal}{Phys. Rev.}
  \textbf{\bibinfo{volume}{D73}}, \bibinfo{pages}{112001}
  (\bibinfo{year}{2006}), \eprint{hep-ex/0508053}.

\bibitem{Ahmad:2002jz}
\bibinfo{author}{\bibfnamefont{Q.~R.} \bibnamefont{Ahmad}} \emph{et~al.}
  (\bibinfo{collaboration}{SNO}), \bibinfo{journal}{Phys. Rev. Lett.}
  \textbf{\bibinfo{volume}{89}}, \bibinfo{pages}{011301}
  (\bibinfo{year}{2002}), \eprint[http://arXiv.org/abs]{nucl-ex/0204008}.

\bibitem{Aharmim:2008kc}
\bibinfo{author}{\bibfnamefont{B.}~\bibnamefont{Aharmim}} \emph{et~al.}
  (\bibinfo{collaboration}{SNO}), \bibinfo{journal}{Phys. Rev. Lett.}
  \textbf{\bibinfo{volume}{101}}, \bibinfo{pages}{111301}
  (\bibinfo{year}{2008}), \eprint{0806.0989}.

\bibitem{Arpesella:2008mt}
\bibinfo{author}{\bibfnamefont{C.}~\bibnamefont{Arpesella}} \emph{et~al.}
  (\bibinfo{collaboration}{Borexino}), \bibinfo{journal}{Phys. Rev. Lett.}
  \textbf{\bibinfo{volume}{101}}, \bibinfo{pages}{091302}
  (\bibinfo{year}{2008}), \eprint{0805.3843}.

\bibitem{Fukuda:1998mi}
\bibinfo{author}{\bibfnamefont{Y.}~\bibnamefont{Fukuda}} \emph{et~al.}
  (\bibinfo{collaboration}{Super-Kamiokande}), \bibinfo{journal}{Phys. Rev.
  Lett.} \textbf{\bibinfo{volume}{81}}, \bibinfo{pages}{1562}
  (\bibinfo{year}{1998}), \eprint{hep-ex/9807003}.

\bibitem{Ashie:2005ik}
\bibinfo{author}{\bibfnamefont{Y.}~\bibnamefont{Ashie}} \emph{et~al.}
  (\bibinfo{collaboration}{Super-Kamiokande}), \bibinfo{journal}{Phys. Rev.}
  \textbf{\bibinfo{volume}{D71}}, \bibinfo{pages}{112005}
  (\bibinfo{year}{2005}), \eprint{hep-ex/0501064}.

\bibitem{Araki:2004mb}
\bibinfo{author}{\bibfnamefont{T.}~\bibnamefont{Araki}} \emph{et~al.}
  (\bibinfo{collaboration}{KamLAND}), \bibinfo{journal}{Phys. Rev. Lett.}
  \textbf{\bibinfo{volume}{94}}, \bibinfo{pages}{081801}
  (\bibinfo{year}{2005}), \eprint{hep-ex/0406035}.

\bibitem{:2008ee}
\bibinfo{author}{\bibfnamefont{S.}~\bibnamefont{Abe}} \emph{et~al.}
  (\bibinfo{collaboration}{KamLAND}), \bibinfo{journal}{Phys. Rev. Lett.}
  \textbf{\bibinfo{volume}{100}}, \bibinfo{pages}{221803}
  (\bibinfo{year}{2008}), \eprint{0801.4589}.

\bibitem{Ahn:2006zz}
\bibinfo{author}{\bibfnamefont{M.~H.} \bibnamefont{Ahn}} \emph{et~al.}
  (\bibinfo{collaboration}{K2K}), \bibinfo{journal}{Phys. Rev.}
  \textbf{\bibinfo{volume}{D74}}, \bibinfo{pages}{072003}
  (\bibinfo{year}{2006}), \eprint{hep-ex/0606032}.

\bibitem{Adamson:2008zt}
\bibinfo{author}{\bibfnamefont{P.}~\bibnamefont{Adamson}} \emph{et~al.}
  (\bibinfo{collaboration}{MINOS}), \bibinfo{journal}{Phys. Rev. Lett.}
  \textbf{\bibinfo{volume}{101}}, \bibinfo{pages}{131802}
  (\bibinfo{year}{2008}), \eprint{0806.2237}.

\bibitem{Ardellier:2004ui}
\bibinfo{author}{\bibfnamefont{F.}~\bibnamefont{Ardellier}} \emph{et~al.}
  (\bibinfo{year}{2004}), \eprint{hep-ex/0405032}.

\bibitem{Guo:2007ug}
\bibinfo{author}{\bibfnamefont{X.}~\bibnamefont{Guo}} \emph{et~al.}
  (\bibinfo{collaboration}{Daya-Bay})  (\bibinfo{year}{2007}),
  \eprint{hep-ex/0701029}.

\bibitem{Kim:2008zzb}
\bibinfo{author}{\bibfnamefont{S.-B.} \bibnamefont{Kim}}
  (\bibinfo{collaboration}{RENO}), \bibinfo{journal}{AIP Conf. Proc.}
  \textbf{\bibinfo{volume}{981}}, \bibinfo{pages}{205} (\bibinfo{year}{2008}).

\bibitem{Itow:2001ee}
\bibinfo{author}{\bibfnamefont{Y.}~\bibnamefont{Itow}} \emph{et~al.},
  \bibinfo{journal}{Nucl. Phys. Proc. Suppl.} \textbf{\bibinfo{volume}{111}},
  \bibinfo{pages}{146} (\bibinfo{year}{2001}),
  \eprint[http://arXiv.org/abs]{hep-ex/0106019}.

\bibitem{Ambats:2004js}
\bibinfo{author}{\bibfnamefont{I.}~\bibnamefont{Ambats}} \emph{et~al.}
  (\bibinfo{collaboration}{NOvA})  (\bibinfo{year}{2004}),
  \eprint{hep-ex/0503053}.

\bibitem{Minakata:2002jv}
\bibinfo{author}{\bibfnamefont{H.}~\bibnamefont{Minakata}},
  \bibinfo{author}{\bibfnamefont{H.}~\bibnamefont{Sugiyama}},
  \bibinfo{author}{\bibfnamefont{O.}~\bibnamefont{Yasuda}},
  \bibinfo{author}{\bibfnamefont{K.}~\bibnamefont{Inoue}}, \bibnamefont{and}
  \bibinfo{author}{\bibfnamefont{F.}~\bibnamefont{Suekane}},
  \bibinfo{journal}{Phys. Rev.} \textbf{\bibinfo{volume}{D68}},
  \bibinfo{pages}{033017} (\bibinfo{year}{2003}), \eprint{hep-ph/0211111}.

\bibitem{Huber:2002rs}
\bibinfo{author}{\bibfnamefont{P.}~\bibnamefont{Huber}},
  \bibinfo{author}{\bibfnamefont{M.}~\bibnamefont{Lindner}}, \bibnamefont{and}
  \bibinfo{author}{\bibfnamefont{W.}~\bibnamefont{Winter}},
  \bibinfo{journal}{Nucl. Phys.} \textbf{\bibinfo{volume}{B654}},
  \bibinfo{pages}{3} (\bibinfo{year}{2003}), \eprint{hep-ph/0211300}.

\bibitem{Huber:2003pm}
\bibinfo{author}{\bibfnamefont{P.}~\bibnamefont{Huber}},
  \bibinfo{author}{\bibfnamefont{M.}~\bibnamefont{Lindner}},
  \bibinfo{author}{\bibfnamefont{T.}~\bibnamefont{Schwetz}}, \bibnamefont{and}
  \bibinfo{author}{\bibfnamefont{W.}~\bibnamefont{Winter}},
  \bibinfo{journal}{Nucl. Phys.} \textbf{\bibinfo{volume}{B665}},
  \bibinfo{pages}{487} (\bibinfo{year}{2003}), \eprint{hep-ph/0303232}.

\bibitem{Huber:2004ug}
\bibinfo{author}{\bibfnamefont{P.}~\bibnamefont{Huber}},
  \bibinfo{author}{\bibfnamefont{M.}~\bibnamefont{Lindner}},
  \bibinfo{author}{\bibfnamefont{M.}~\bibnamefont{Rolinec}},
  \bibinfo{author}{\bibfnamefont{T.}~\bibnamefont{Schwetz}}, \bibnamefont{and}
  \bibinfo{author}{\bibfnamefont{W.}~\bibnamefont{Winter}},
  \bibinfo{journal}{Phys. Rev.} \textbf{\bibinfo{volume}{D70}},
  \bibinfo{pages}{073014} (\bibinfo{year}{2004}), \eprint{hep-ph/0403068}.

\bibitem{McConnel:2004bd}
\bibinfo{author}{\bibfnamefont{K.~B.} \bibnamefont{McConnel}} \bibnamefont{and}
  \bibinfo{author}{\bibfnamefont{M.~H.} \bibnamefont{Shaevitz}},
  \bibinfo{journal}{Int. J. Mod. Phys.} \textbf{\bibinfo{volume}{A21}},
  \bibinfo{pages}{3825} (\bibinfo{year}{2006}), \eprint{hep-ex/0409028}.

\bibitem{minos-app}
\bibinfo{author}{\bibfnamefont{M.}~\bibnamefont{Sanchez}}
  (\bibinfo{collaboration}{MINOS}), \bibinfo{note}{{Talk given at Fermilab, 27
  Feb. 2009}}.

\bibitem{Rosa:2008zz}
\bibinfo{author}{\bibfnamefont{G.}~\bibnamefont{Rosa}}
  (\bibinfo{collaboration}{OPERA}), \bibinfo{journal}{J. Phys. Conf. Ser.}
  \textbf{\bibinfo{volume}{136}}, \bibinfo{pages}{022015}
  (\bibinfo{year}{2008}).

\bibitem{euronu}
\emph{\bibinfo{title}{{European Commission FP7 Design Study: A High Intensity
  Neutrino Oscillation Facility in Europe}}}, \bibinfo{note}{{\tt
  http://www.euronu.org}}.

\bibitem{ids}
\emph{\bibinfo{title}{International design study of the neutrino factory}},
  \bibinfo{note}{{\tt http://www.ids-nf.org}}.

\bibitem{Huber:2004ka}
\bibinfo{author}{\bibfnamefont{P.}~\bibnamefont{Huber}},
  \bibinfo{author}{\bibfnamefont{M.}~\bibnamefont{Lindner}}, \bibnamefont{and}
  \bibinfo{author}{\bibfnamefont{W.}~\bibnamefont{Winter}},
  \bibinfo{journal}{Comput. Phys. Commun.} \textbf{\bibinfo{volume}{167}},
  \bibinfo{pages}{195} (\bibinfo{year}{2005}), \bibinfo{note}{{\tt
  http://www.mpi-hd.mpg.de/lin/globes/}}, \eprint{hep-ph/0407333}.

\bibitem{Huber:2007ji}
\bibinfo{author}{\bibfnamefont{P.}~\bibnamefont{Huber}},
  \bibinfo{author}{\bibfnamefont{J.}~\bibnamefont{Kopp}},
  \bibinfo{author}{\bibfnamefont{M.}~\bibnamefont{Lindner}},
  \bibinfo{author}{\bibfnamefont{M.}~\bibnamefont{Rolinec}}, \bibnamefont{and}
  \bibinfo{author}{\bibfnamefont{W.}~\bibnamefont{Winter}},
  \bibinfo{journal}{Comput. Phys. Commun.} \textbf{\bibinfo{volume}{177}},
  \bibinfo{pages}{432} (\bibinfo{year}{2007}), \eprint{hep-ph/0701187}.

\bibitem{Huber:2006vr}
\bibinfo{author}{\bibfnamefont{P.}~\bibnamefont{Huber}},
  \bibinfo{author}{\bibfnamefont{J.}~\bibnamefont{Kopp}},
  \bibinfo{author}{\bibfnamefont{M.}~\bibnamefont{Lindner}},
  \bibinfo{author}{\bibfnamefont{M.}~\bibnamefont{Rolinec}}, \bibnamefont{and}
  \bibinfo{author}{\bibfnamefont{W.}~\bibnamefont{Winter}},
  \bibinfo{journal}{JHEP} \textbf{\bibinfo{volume}{05}}, \bibinfo{pages}{072}
  (\bibinfo{year}{2006}), \eprint{hep-ph/0601266}.

\bibitem{Mention:2007um}
\bibinfo{author}{\bibfnamefont{G.}~\bibnamefont{Mention}},
  \bibinfo{author}{\bibfnamefont{T.}~\bibnamefont{Lasserre}}, \bibnamefont{and}
  \bibinfo{author}{\bibfnamefont{D.}~\bibnamefont{Motta}}
  (\bibinfo{year}{2007}), \eprint{0704.0498}.

\bibitem{Fechner2006}
\bibinfo{author}{\bibfnamefont{M.}~\bibnamefont{Fechner}},
  \emph{\bibinfo{title}{D\'etermination des performances attendues sur la
  recherche de l'oscillation $\nu_\mu\rightarrow\nu_e$ dans l'experi\'ence T2K
  depuis l'\'etude des donn\'ees recueilles dans l'\'experience K2K}}, Ph.D.
  thesis, \bibinfo{school}{Universit\'e Paris VI} (\bibinfo{year}{2006}).

\bibitem{Kato:2008zz}
\bibinfo{author}{\bibfnamefont{I.}~\bibnamefont{Kato}}
  (\bibinfo{collaboration}{T2K}), \bibinfo{journal}{J. Phys. Conf. Ser.}
  \textbf{\bibinfo{volume}{136}}, \bibinfo{pages}{022018}
  (\bibinfo{year}{2008}).

\bibitem{Yang_2004}
\bibinfo{author}{\bibfnamefont{T.}~\bibnamefont{Yang}} \bibnamefont{and}
  \bibinfo{author}{\bibfnamefont{S.}~\bibnamefont{Woijcicki}}
  (\bibinfo{collaboration}{NOvA})  (\bibinfo{year}{2004}),
  \eprint{Off-Axis-Note-SIM-30}.

\bibitem{nova_TDR}
\emph{\bibinfo{title}{{NOvA Technical Design Report, October 23, 2007}}},
  \bibinfo{note}{{\tt
  http://www-nova.fnal.gov/nova\_cd2\_review/tdr\_oct\_23/tdr.htm}}.

\bibitem{Schwetz:2008er}
\bibinfo{author}{\bibfnamefont{T.}~\bibnamefont{Schwetz}},
  \bibinfo{author}{\bibfnamefont{M.}~\bibnamefont{Tortola}}, \bibnamefont{and}
  \bibinfo{author}{\bibfnamefont{J.~W.~F.} \bibnamefont{Valle}},
  \bibinfo{journal}{New J. Phys.} \textbf{\bibinfo{volume}{10}},
  \bibinfo{pages}{113011} (\bibinfo{year}{2008}), \eprint{0808.2016}.

\bibitem{McFarlane:2009xx}
\bibinfo{author}{\bibfnamefont{M.}~\bibnamefont{McFarlane}},
  \bibinfo{author}{\bibfnamefont{K.}~\bibnamefont{Heeger}},
  \bibinfo{author}{\bibfnamefont{P.}~\bibnamefont{Huber}},
  \bibinfo{author}{\bibfnamefont{C.}~\bibnamefont{Lewis}}, \bibnamefont{and}
  \bibinfo{author}{\bibfnamefont{W.}~\bibnamefont{Wang}}
  (\bibinfo{year}{2009}), \bibinfo{note}{in preperation}.

\bibitem{Fogli:1996pv}
\bibinfo{author}{\bibfnamefont{G.~L.} \bibnamefont{Fogli}} \bibnamefont{and}
  \bibinfo{author}{\bibfnamefont{E.}~\bibnamefont{Lisi}},
  \bibinfo{journal}{Phys. Rev.} \textbf{\bibinfo{volume}{D54}},
  \bibinfo{pages}{3667} (\bibinfo{year}{1996}), \eprint{hep-ph/9604415}.

\bibitem{Antusch:2004yx}
\bibinfo{author}{\bibfnamefont{S.}~\bibnamefont{Antusch}},
  \bibinfo{author}{\bibfnamefont{P.}~\bibnamefont{Huber}},
  \bibinfo{author}{\bibfnamefont{J.}~\bibnamefont{Kersten}},
  \bibinfo{author}{\bibfnamefont{T.}~\bibnamefont{Schwetz}}, \bibnamefont{and}
  \bibinfo{author}{\bibfnamefont{W.}~\bibnamefont{Winter}},
  \bibinfo{journal}{Phys. Rev.} \textbf{\bibinfo{volume}{D70}},
  \bibinfo{pages}{097302} (\bibinfo{year}{2004}), \eprint{hep-ph/0404268}.

\bibitem{Niehage:2008sg}
\bibinfo{author}{\bibfnamefont{S.}~\bibnamefont{Niehage}} \bibnamefont{and}
  \bibinfo{author}{\bibfnamefont{W.}~\bibnamefont{Winter}},
  \bibinfo{journal}{Phys. Rev.} \textbf{\bibinfo{volume}{D78}},
  \bibinfo{pages}{013007} (\bibinfo{year}{2008}), \eprint{0804.1546}.

\bibitem{Antusch:2003kp}
\bibinfo{author}{\bibfnamefont{S.}~\bibnamefont{Antusch}},
  \bibinfo{author}{\bibfnamefont{J.}~\bibnamefont{Kersten}},
  \bibinfo{author}{\bibfnamefont{M.}~\bibnamefont{Lindner}}, \bibnamefont{and}
  \bibinfo{author}{\bibfnamefont{M.}~\bibnamefont{Ratz}},
  \bibinfo{journal}{Nucl. Phys.} \textbf{\bibinfo{volume}{B674}},
  \bibinfo{pages}{401} (\bibinfo{year}{2003}), \eprint{hep-ph/0305273}.

\bibitem{REFDC}
\bibinfo{author}{\bibfnamefont{S.}~\bibnamefont{Peeters}},
  \emph{\bibinfo{title}{Seeking $\theta_{13}$ with reactor neutrinos}},
  \bibinfo{note}{{Talk given at NOW 2008 {\tt
  http://www.ba.infn.it/$\sim$now/now2008/}}}.

\bibitem{REFDYB}
\bibinfo{author}{\bibfnamefont{R.}~\bibnamefont{McKeown}},
  \emph{\bibinfo{title}{The daya bay reactor neutrino experiment}},
  \bibinfo{note}{{Talk given at CIPANP 2009, San Diego, USA {\tt
  http://groups.physics.umn.edu/cipanp2009/}}}.

\bibitem{REFRENO}
\bibinfo{author}{\bibfnamefont{Y.}~\bibnamefont{Oh}},
  \emph{\bibinfo{title}{{Current status of RENO}}}, \bibinfo{note}{{Talk given
  at NOW 2008 \tt http://www.ba.infn.it/$\sim$now/now2008/}}.

\bibitem{REFT2K}
\bibinfo{author}{\bibfnamefont{H.}~\bibnamefont{Kakuno}},
  \emph{\bibinfo{title}{{Neutrino oscillations at T2K}}}, \bibinfo{note}{{Talk
  given at NOW 2008 \tt http://www.ba.infn.it/$\sim$now/now2008/}}.

\bibitem{REFNOVA}
\bibinfo{author}{\bibfnamefont{M.}~\bibnamefont{Messier}},
  \emph{\bibinfo{title}{{The NOvA Experiment at Fermilab}}},
  \bibinfo{note}{{Talk given at ICHEP 2008, \tt
  http://www.hep.upenn.edu/ichep08/talks/misc/schedule}}.

\bibitem{Schwetz:2005fy}
\bibinfo{author}{\bibfnamefont{T.}~\bibnamefont{Schwetz}} \bibnamefont{and}
  \bibinfo{author}{\bibfnamefont{W.}~\bibnamefont{Winter}},
  \bibinfo{journal}{Phys. Lett.} \textbf{\bibinfo{volume}{B633}},
  \bibinfo{pages}{557} (\bibinfo{year}{2006}), \eprint{hep-ph/0511177}.

\bibitem{Kopp:2007ne}
\bibinfo{author}{\bibfnamefont{J.}~\bibnamefont{Kopp}},
  \bibinfo{author}{\bibfnamefont{M.}~\bibnamefont{Lindner}},
  \bibinfo{author}{\bibfnamefont{T.}~\bibnamefont{Ota}}, \bibnamefont{and}
  \bibinfo{author}{\bibfnamefont{J.}~\bibnamefont{Sato}},
  \bibinfo{journal}{Phys. Rev.} \textbf{\bibinfo{volume}{D77}},
  \bibinfo{pages}{013007} (\bibinfo{year}{2008}), \eprint{0708.0152}.

\bibitem{NAKAYA}
\bibinfo{author}{\bibfnamefont{T.}~\bibnamefont{Nakaya}},
  \bibinfo{note}{private communication}.

\bibitem{T2KUPGRADE}
\bibinfo{author}{\bibfnamefont{K.}~\bibnamefont{Hasegawa}},
  \emph{\bibinfo{title}{The j-parc neutrino beam}}, \bibinfo{note}{{Talk given
  at NNN 08, Paris France, {\tt http://nnn08.in2p3.fr/} }}.

\bibitem{PROJECTX}
\emph{\bibinfo{title}{Resource-loaded schedule}}, \bibinfo{note}{director's
  Preliminary Cost and Schedule Review, March 16-17, 2009, {\tt
  http://projectx.fnal.gov}}.

\bibitem{Minakata:2003wq}
\bibinfo{author}{\bibfnamefont{H.}~\bibnamefont{Minakata}} \bibnamefont{and}
  \bibinfo{author}{\bibfnamefont{H.}~\bibnamefont{Sugiyama}},
  \bibinfo{journal}{Phys. Lett.} \textbf{\bibinfo{volume}{B580}},
  \bibinfo{pages}{216} (\bibinfo{year}{2004}), \eprint{hep-ph/0309323}.

\bibitem{Winter:2003ye}
\bibinfo{author}{\bibfnamefont{W.}~\bibnamefont{Winter}},
  \bibinfo{journal}{Phys. Rev.} \textbf{\bibinfo{volume}{D70}},
  \bibinfo{pages}{033006} (\bibinfo{year}{2004}), \eprint{hep-ph/0310307}.

\end{thebibliography}

\end{document}